\newif\ifconfver
\confverfalse      
\confvertrue        

\newif\ifplainver  
\plainvertrue

\newif\ifhide  
\hidetrue

\ifplainver
    \confverfalse   
\fi

\ifconfver
     \documentclass[10pt,twocolumn,twoside]{IEEEtran}
\else
    \ifplainver
        \documentclass[11pt]{article}
        \usepackage{fullpage}
    \else
        \documentclass[12pt,draftcls,onecolumn]{IEEEtran}
    \fi
\fi

\usepackage{etoolbox,soul}%
\usepackage{xpatch}
\usepackage{blindtext}
\usepackage{tocloft}%
\newlength{\articlesectionshift}%
\let\LaTeXStandardSection\section
\let\LaTeXStandardTheSection\thesection

\let\LaTeXStandardTheSubSubSection\thesubsubsection
\let\LaTeXStandardTheParagraph\theparagraph

\makeatletter
\newcounter{titlecounter}

\xpretocmd{\maketitle}{\ifnumgreater{\value{titlecounter}}{1}}{\clearpage}{}{} 
\xpatchcmd{\maketitle}{\let\maketitle\relax\let\@maketitle\relax}{\refstepcounter{titlecounter}\begingroup
  \addtocontents{toc}{\begingroup\addtolength{\cftsecindent}{-\articlesectionshift}}%
  \addcontentsline{toc}{section}{\protect{\numberline{\thetitlecounter}{\@title~ \@author}}}%
  \addtocontents{toc}{\endgroup}
}{%
  \typeout{Patching was successful}
}{%
  \typeout{patching failed}
}%

\def\@IEEEdestroythesectionargument#1{\LaTeXStandardSection{#1}}%

\xapptocmd{\maketitle}{%
\renewcommand{\thesection}{\LaTeXStandardTheSection}%
\renewcommand{\thesubsection}{\LaTeXStandardTheSubSection}%
\renewcommand{\thesubsubsection}{\LaTeXStandardTheSubSubSection}%
\renewcommand{\theparagraph}{\LaTeXStandardTheParagraph}%
}{}{}%

\@addtoreset{section}{titlecounter}

\usepackage{calc,amsfonts,amssymb,amsmath,bm,url,color,theorem,graphicx,cite}
\usepackage{psfrag,float}
\usepackage{algorithm}
\usepackage{algpseudocode}
\usepackage{ marvosym }
\usepackage{soul}
\usepackage{enumerate}
\usepackage{bbm}
\usepackage{shortcuts_OPT}
\usepackage{multirow}
\usepackage{titling}

\usepackage[labelformat=simple]{subcaption}

\usepackage{eqparbox}

\definecolor{orange}{RGB}{255,107,0}
\def\blue{\color{blue}}
\def\red{\color{red}}

\newtheorem{Fact}{Fact}
\newtheorem{Lemma}{Lemma}
\newtheorem{Prop}{Proposition}
\newtheorem{Theorem}{Theorem}

\newtheorem{Asm}{Assumption}
\theorembodyfont{\rmfamily}

\newtheorem{Remark}{Remark}

\newcommand\SCA{{simplex component analysis}}
\newcommand\CapSCA{{Simplex component analysis}}

\newcommand\bw{\ensuremath{{\bm w}}}

\newcommand\bx{\ensuremath{{\bm x}}}
\newcommand\by{\ensuremath{{\bm y}}}
\newcommand\bG{\ensuremath{{\bm G}}}

\newcommand\be{\ensuremath{{\bm e}}}
\newcommand\bz{\ensuremath{{\bm z}}}

\newcommand\bR{\ensuremath{{\bm R}}}

\newcommand\bX{\ensuremath{{\bm X}}}

\newcommand\bC{\ensuremath{{\bm C}}}
\newcommand\bc{\ensuremath{{\bm c}}}
\newcommand\ba{\ensuremath{{\bm a}}}

\newcommand\bA{\ensuremath{{\bm A}}}

\newcommand\bb{\ensuremath{{\bm b}}}

\newcommand\bB{\ensuremath{{\bm B}}}
\newcommand\blam{\ensuremath{{\bm \lambda}}}
\newcommand\bmu{\ensuremath{{\bm \mu}}}
\newcommand\balp{\ensuremath{{\bm \alpha}}}
\newcommand\bbeta{\ensuremath{{\bm \beta}}}
\newcommand\bXi{\ensuremath{{\bm \Xi}}}
\newcommand\bPi{\ensuremath{{\bm \Pi}}}

\newcommand\bF{\ensuremath{{\bm F}}}

\newcommand\bd{\ensuremath{{\bm d}}}

\newcommand\bv{\ensuremath{{\bm v}}}

\newcommand\bTheta{\ensuremath{{\bm \Theta}}}

\newcommand\bxi{\ensuremath{{\bm \xi}}}
\newcommand\bzeta{\ensuremath{{\bm \zeta}}}

\newcommand\bY{\ensuremath{{\bm Y}}}

\newcommand\bU{\ensuremath{{\bm U}}}
\newcommand\bs{\ensuremath{{\bm s}}}
\newcommand\bS{\ensuremath{{\bm S}}}

\newcommand{\Rbb}{\mathbb{R}}

\newcommand{\setA}{\mathcal{A}}
\newcommand{\setB}{\mathcal{B}}
\newcommand{\setD}{\mathcal{D}}
\newcommand{\setH}{\mathcal{H}}
\newcommand{\setX}{\mathcal{X}}

\newcommand{\setC}{\mathcal{C}}
\newcommand{\setN}{\mathcal{N}}

\newcommand{\Exp}{\mathbb{E}}

\newcommand{\Diag}{\mathrm{Diag}}
\newcommand{\diag}{\mathrm{diag}}

\newcommand\bQ{\ensuremath{{\bm Q}}}

\newcommand{\bzero}{{\bm 0}}
\newcommand{\bone}{{\bm 1}}
\newcommand{\bI}{{\bm I}}

\newcommand\indfn[1]{{{\mathbbm 1}_{#1}}}

\newcommand\sspan{\ensuremath{{\rm span}}}
\newcommand\aff{\ensuremath{{\rm aff}}}
\newcommand\conv{\ensuremath{{\rm conv}}}
\newcommand\bconv{\ensuremath{\overline{\rm conv}}}
\newcommand\var{\ensuremath{{\rm var}}}
\newcommand\cov{\ensuremath{{\rm Cov}}}
\newcommand\svol{\ensuremath{{\rm vol}}}
\newcommand\TVar{\ensuremath{{\rm tvar}}}
\newcommand\tr{\ensuremath{{\rm tr}}}

\usepackage{tikz}
\usetikzlibrary{arrows.meta}
\usetikzlibrary{decorations}
\usetikzlibrary{calc}
\usetikzlibrary{shapes.geometric}
\usetikzlibrary{external}

\begin{document}


\newcommand{\papertitle}{
Probabilistic Simplex Component Analysis
}

\newcommand{\paperabstract}{
This study presents PRISM, a probabilistic simplex component analysis approach to identifying the vertices of a data-circumscribing simplex from data.
The problem has a rich variety of applications,
the most notable being hyperspectral unmixing in remote sensing and non-negative matrix factorization in machine learning.
PRISM uses a simple probabilistic model, namely, uniform simplex data distribution and additive Gaussian noise,
and it carries out inference by maximum likelihood.
The inference model is sound in the sense that the vertices are provably identifiable under some assumptions,
and it suggests that PRISM can be effective in combating noise when the number of data points is large.
PRISM has strong, but hidden, relationships with simplex volume minimization, a powerful geometric approach for the same problem.
We study these fundamental aspects,
and we also consider algorithmic schemes based on importance sampling and variational inference.
In particular, the variational inference scheme is shown to resemble a matrix factorization problem with a special regularizer, which draws an interesting connection to the matrix factorization approach. 
Numerical results are provided to demonstrate the potential of PRISM.
}


\ifplainver


\title{\papertitle}

\author{
	Ruiyuan Wu$^\dag$, Wing-Kin Ma$^\dag$, Yuening Li$^\dag$, Anthony Man-Cho  So$^\ddag$, \\ and Nicholas D. Sidiropoulos$^\star$ \\ ~ \\ 
	$^\dag$Department of Electronic Engineering, The Chinese University of Hong Kong, \\
	Hong Kong SAR of China \\ ~ \\
	$^\ddag$Department of Systems Engineering and Engineering Management, \\ The Chinese University of Hong Kong, 
	Hong Kong SAR of China \\ ~ \\
	$^\star$Department of Electrical and Computer Engineering, \\
	University of Virginia, Charlottesville, Virginia 22904, USA	
}

\maketitle

\begin{abstract}
	\paperabstract
\end{abstract}

\bigskip


\else
\title{\papertitle}

\ifconfver \else {\linespread{1.1} \rm \fi
	
	\author{
		Ruiyuan Wu, Wing-Kin Ma, Yuening Li, Anthony  Man-Cho So, and Nicholas D. Sidiropoulos 
	}

	\maketitle
	
	\ifconfver \else
	\begin{center} \vspace*{-2\baselineskip}
	\end{center}
	\fi
	
	\begin{abstract}
		\paperabstract
	\end{abstract}
	
	
	\begin{IEEEkeywords}\vspace{-0.0cm}
		Simplex-structured matrix factorization, maximum
		likelihood, identifiability,  simplex volume minimization, variational inference, 
		hyperspectral unmixing
	\end{IEEEkeywords}
	
	\ifconfver \else \IEEEpeerreviewmaketitle} \fi

\fi

\ifconfver \else
\ifplainver \else
\newpage
\fi \fi


\newpage 

~

\vfill

\noindent 
\section*{Acknowledgments}
This work is dedicated to the late professor Jos\'{e} Bioucas-Dias.
Wing-Kin Ma is greatly indebted to him for his many inspirations, support, encouragement, and occasionally hard time in the form of thought-provoking challenges; the same goes for the many wonderful interactions with him over a decade.
Had they not met, this work could have never existed.

\vfill

~

\newpage


\section{Introduction}

\begin{figure}[h]
	\centering
	\includegraphics[width=.33\textwidth]{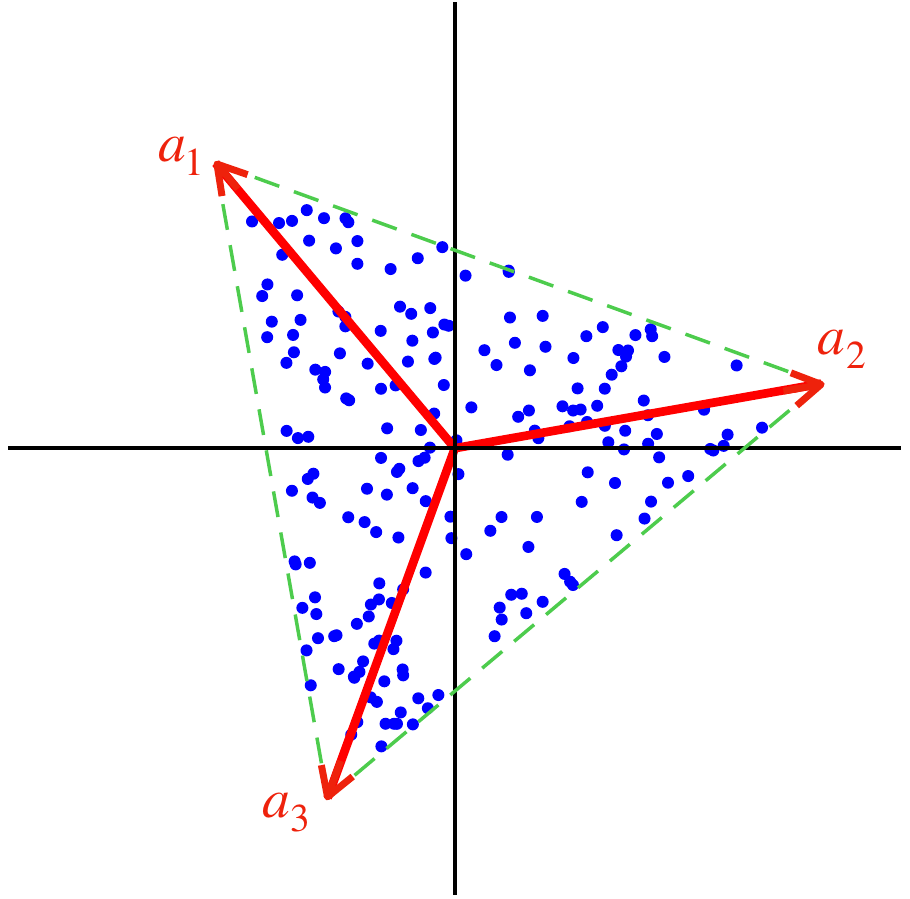}
	\caption{Illustration of \SCA\ by a scatterplot. Blue dots: data points $\by_t$'s,  green line: the data circumscribing simplex, red arrows: vertices $\ba_i$'s of the data circumscribing simplex.}
	\label{fig:scatterplots}
\end{figure}

Consider this problem:
We have a collection of multi-dimensional data points that are circumscribed by a simplex;
see Fig.~\ref{fig:scatterplots} for an illustration.
Can we learn the vertices of that simplex from the data points?
The pursuit of a solution to such vertex-finding problem is termed {\em \SCA} in this study.
In particular, we will consider a probabilistic \SCA\ approach that employs the same inference formulation as probabilistic principal component analysis (PCA)~\cite{tipping1999probabilistic} and independent component analysis (ICA)~\cite{pham1997blind,attias1999independent,khemakhem2020variational}.

\subsection{Background and State of the Art}

\CapSCA\ arose in different fields and has many names.
It appears in hyperspectral unmixing (HU) in remote sensing \cite{Jose12,Ma2014HU}, a topic that has more than 30 years of history \cite{craig1990unsupervised}.
It emerges in non-negative matrix factorization (NMF) in machine learning \cite{arora2016computing,gillis2014,fu2019nonnegative}, with application to topic modeling.
Some other areas also stumbled on the same problem; see, e.g., \cite{Ma2014HU,gillis2014,fu2019nonnegative} and the references therein, for details.
It is now recognized that \SCA\ covers a rich variety of applications---in addition to the aforementioned HU and topic modeling applications, it has been applied to biomedical imaging \cite{chen2011tissue}, blind audio source separation \cite{Fu2015}, finding representatives from data in computer vision \cite{Elhamifar2012}, community detection \cite{panov2017consistent,pmlr-v97-huang19c}, and crowdsourcing \cite{NIPS2019_8999}, to name a few.

\begin{figure}[h]
	\centering
	\begin{subfigure}[b]{0.33\linewidth}
		\includegraphics[width=\textwidth]{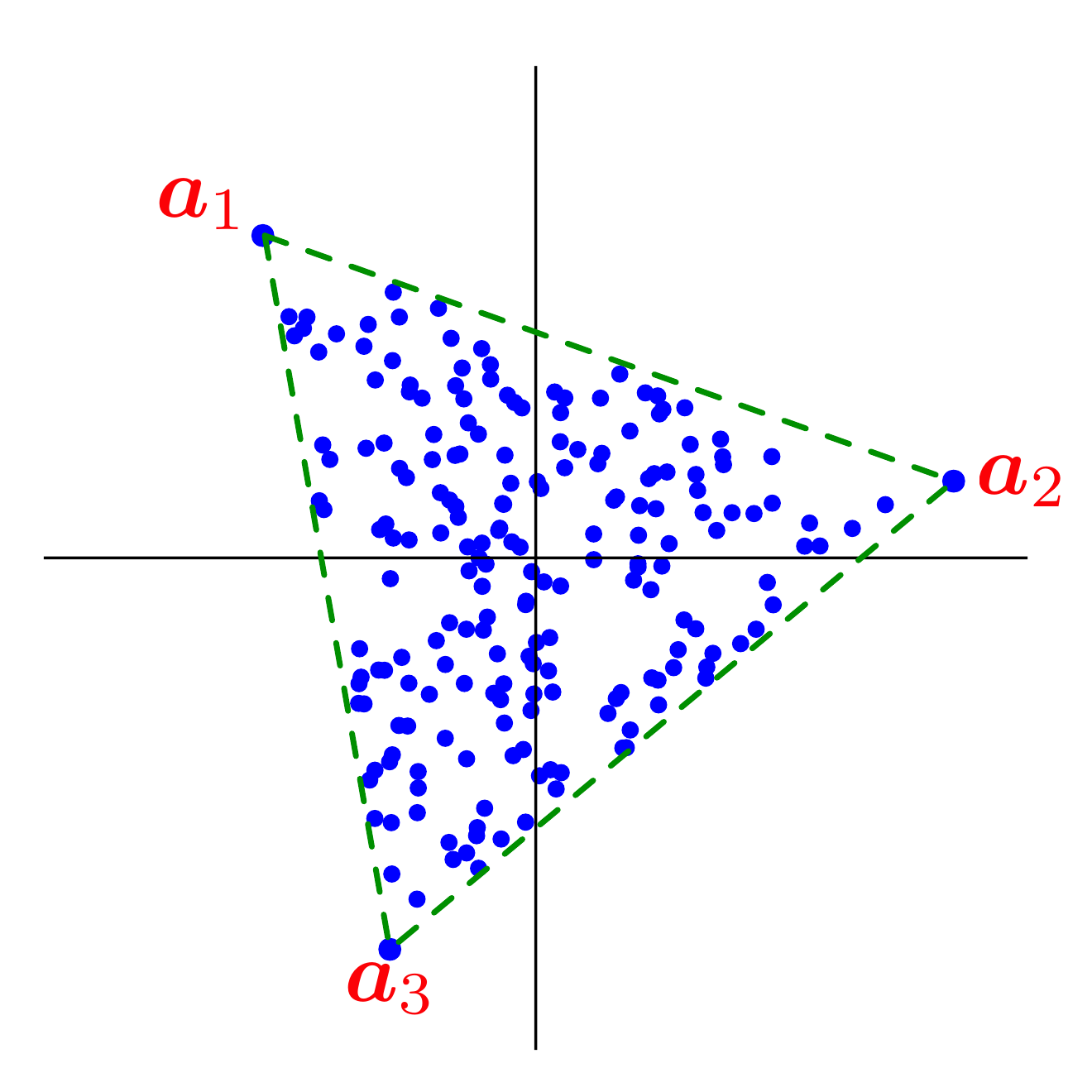}
		\caption{pure-pixel search}
	\end{subfigure}
	\hfil
	\begin{subfigure}[b]{0.33\linewidth}
		\includegraphics[width=\textwidth]{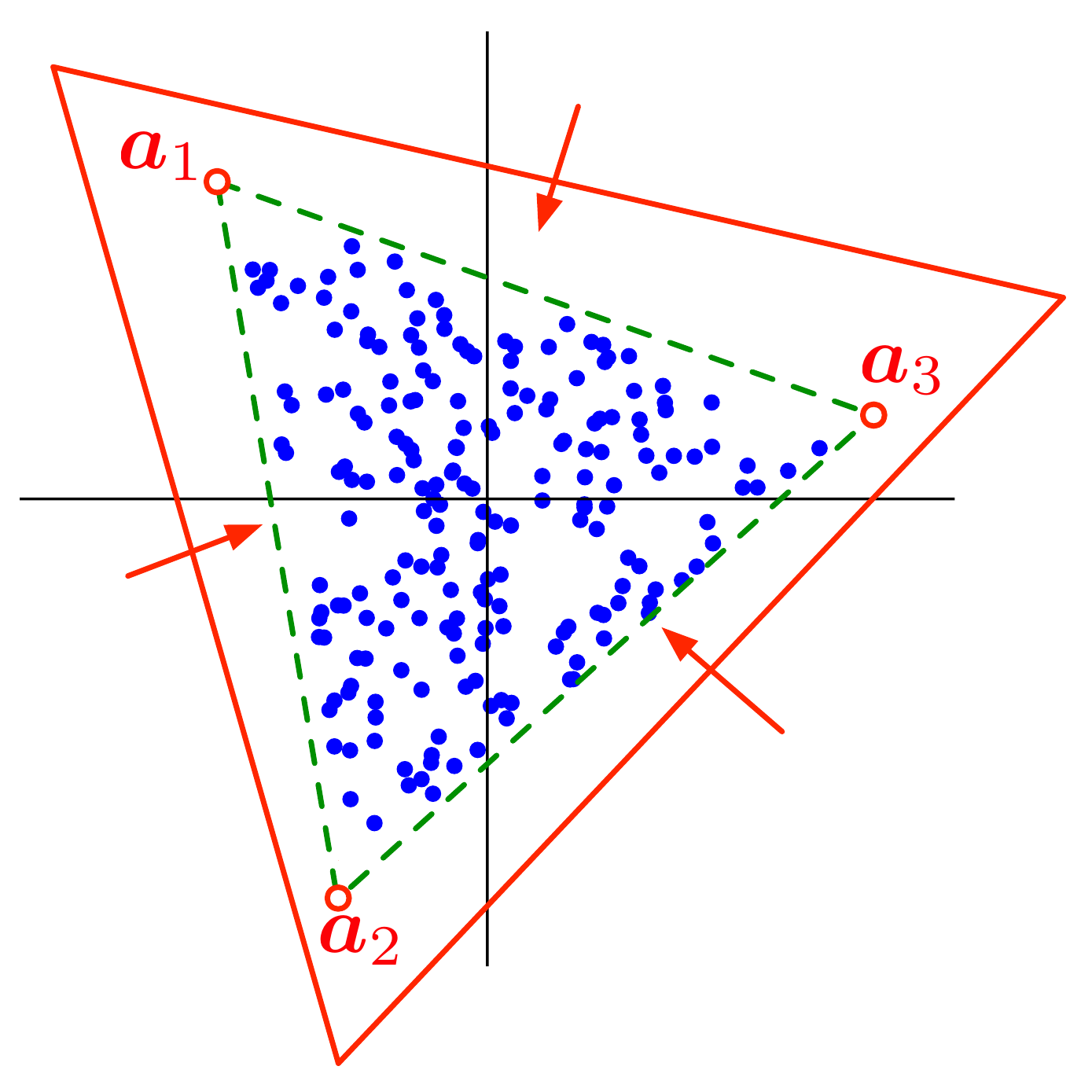}
		\caption{SVMin}
	\end{subfigure}
	\caption{Illustration of the CG concepts. Blue dots: data points $\by_t$'s, green line: the true data circumscribing simplex, red line: a data enclosing simplex.}
	\label{fig:CG}
\end{figure}

In solving the \SCA\ problem, the majority of the existing studies follow, or turn out to be related to, the notion of {\em convex geometry} (CG).
Its idea is to exploit certain geometric structures with the data points.
There are two main approaches with CG.
One, called {\em pure-pixel search} in HU or {\em separable NMF} in machine learning, assumes that some data points lie exactly at the vertices; see Fig.~\ref{fig:CG}(a) for an illustration.
The problem is then to identify those vertex points algorithmically;
see \cite{Winter1999,Nascimento2005,Chan2011,Elhamifar2012,arora2016computing,AGH12,recht2012factoring,gillis2014fast,Ma2014HU,gillis2014,fu2019nonnegative} for the many different ways to do so.
Another, called {\em simplex volume minimization} (SVMin), amounts to finding a simplex that encloses the data points and yields the minimum
volume \cite{Craig1994,Li2008,Chan2009}.
As visualized in Fig.~\ref{fig:CG}(b), the minimum volume data-enclosing simplex seems to coincide with the true data simplex when the data points are adequately well-spread on the simplex.
This intuition has recently been confirmed to be mathematically sound---SVMin can identify the vertices
under a geometric assumption that is likely to hold for sufficiently well-spread data points \cite{lin2015identifiability,Fu2015,fu2016robust}.
That geometric assumption is much more relaxed than the assumption of having vertex points in pure-pixel search or separable NMF,
and thus SVMin is arguably more powerful.
We also refer the reader to \cite{ge2015intersecting,lin2018maximum} for a few other original geometric approaches.

While the notions of CG are elegant, they were established under the noiseless case---at least in the beginning of most of the developments.
In the noisy case some researchers developed  ``provably good'' schemes by equipping their CG algorithms with recovery accuracy analyses, typically under the separable NMF approach;
see, e.g., \cite{arora2016computing,AGH12,recht2012factoring,gillis2014fast}.
Such analyses are fundamentally intriguing in pinning down the noise robustness of separable NMF.
Others altered the formulations 
to make the solutions more robust against noise in practice, and this is more often seen for SVMin; see, e.g., \cite{Chan2011,miao2007endmember,Dias2009,Arul2011,fu2016robust}.
Such alternations usually introduce new parameters, typically for regularization.  
Those parameters are usually tuned in a manual fashion, with no strong theory to guide.

Probabilistic approaches, such as Bayesian and maximum-likelihood (ML) inference,
are arguably more pertinent
when there is noise.
In HU we have seen applications of probabilistic methods to \SCA,
and here we mention two representative developments.
Dobigeon~{\em et.~al}~\cite{dobigeon2009joint} studied Bayesian inference.
The difficulty in that work is that some probability density functions (PDFs) appear as intractable integrals,
and the issue is tackled by Markov chain Monte Carlo (MCMC) sampling which is known to be computationally expensive for large problem dimensions.
Nascimento and Bioucas-Dias studied an ML inference scheme called dependent component analysis (DECA) \cite{nascimento2012hyperspectral}.
DECA considers the noiseless case,
and it employs expectation maximization to realize ML.

\subsection{Contributions of the Present Study}

In this study, we are interested in probabilistic \SCA.
Named PRISM (PRobabIlistic SiMplex),
our approach considers ML inference under the model of uniform simplex distribution with the noise-free components and Gaussian distribution with noise.
The ML inference formulation is principally the same as that by DECA, but with a key difference---{\em noise}.
The likelihood function in the noiseless case, or in DECA, has a closed form.
In the noisy case we no longer have this prestige---the likelihood function appears as an integral that has no known analytical expression in general.
This is an obstacle to be overcome or circumvented.

Our study spans theoretical identifiability, drawing connections to the  CG approach, and exploration of algorithmic realizations. 
Our contributions are summarized as follows.

{\em 1) Identifiability: }
Can PRISM correctly identify the vertices, particularly in the noisy case?
Understanding identifiability is vital in confirming whether we are working on a sound inference model, and it has been a key aspect in the CG studies.
We will confirm that the answer is yes in theory, as far as 
we have a large amount of data points (technically, infinite).
We pin down the identifiability result by leveraging insight from a recent paper on ICA identifiability \cite{khemakhem2020variational}.
The main challenge lies in making the proof mathematically rigorous.

{\em  2) Connections with SVMin in CG:} \
We will show that several representative SVMin methods in CG can be derived from PRISM, either in the noiseless case or as approximations in the noisy case.
These relationships, which are not obvious at first sight, enrich our understanding---the deterministic CG and the stochastic  PRISM are not disparate subjects;
rather, they are intimately related.

We are obliged to commend
Bioucas-Dias who informally, but insightfully, mentioned one of the aforementioned relationships in his WHISPERS 2009 oral presentation~\cite{nascimento2009learning}.
Our task entails
consolidating and expanding his initial insight to discover more connections.
It is interesting to note that, coincidentally, Dobigeon {\em et al.} touched on a result similar to Bioucas-Dias' reporting in the same year~\cite[Appendix]{dobigeon2009joint}.

{\em 3) New Algorithmic Schemes:} \
We take inspiration from statistical inference and 
develop two schemes to realize PRISM algorithmically.
The first is importance sampling approximation (ISA) via  Monte Carlo expectation maximization \cite{wei1990monte}.
The second, which is arguably more interesting, is variational inference approximation (VIA) \cite{bishop2006pattern}.
VIA has recently become popular in statistics, data science and machine learning,
and its idea is to approximate the intractable likelihood function by optimization.
We will propose, and custom-derive, a VIA scheme for PRISM.
Also we will reveal a connection between VIA-PRISM and matrix factorization---VIA-PRISM can be seen as an instance of simplex-structured  matrix factorization, with a special regularization not seen in the previous matrix factorization literature.

\subsection{Comparison with Related Studies and Organization}

Let us further elaborate on the similarities and differences between PRISM and DECA.
As mentioned, the key difference is that PRISM and DECA consider the noisy and noiseless cases, respectively.
Another difference is in their respective aims.
In PRISM we aim to understand basic aspects by adopting a plain uniform simplex distribution model.
In DECA the authors want to learn complex phenomena of real-world data by applying a complex Dirichlet mixture (and non-uniform) distribution model.
Furthermore, 
and beyond the scope of \SCA,
the inference formulations of PRISM and DECA fall into the same genre as those in probabilistic PCA \cite{tipping1999probabilistic} and ICA \cite{pham1997blind,attias1999independent,khemakhem2020variational}.\footnote{Specifically, they all employ latent variable models, 
	and each postulates a different latent prior---independent Gaussian for probabilistic PCA, independent non-Gaussian for probabilistic ICA, simplex-uniform for PRISM, and Dirichlet mixture for DECA.}

Since we will consider VIA for PRISM, it is worth noting that VIA was used in related contexts such as ICA \cite{attias1999independent}, latent Dirichlet allocation \cite{blei2003latent} and nonlinear (or deep) ICA  \cite{khemakhem2020variational}.
However, the VIA of PRISM takes a different appearance from those of the previous studies, due to the different model. 
In fact, technically we will need to devise a specialized method to solve the new optimization problem arising from VIA-PRISM.

We should note that most of the results to be presented appear for the first time;
the exception is the results for ISA, which were reported in conferences \cite{wu2017stochastic,wu2019stochastic,li2020stochastic}.
The nature of our study is fundamental, exploring the potential of PRISM as a framework.
We will focus less on computational or implementation aspects, which will be future work.

This article is organized as follows.
Section~\ref{sect:prelim} concisely reviews the necessary concepts to understand this work.
Section~\ref{sect:formulation} describes the  PRISM model and formulation.
Section~\ref{sect:id} studies the PRISM identifiability.
Section~\ref{sect:relate} shows the hidden relationships of PRISM and SVMin.
Sections~\ref{sect:PRISM_isa}--\ref{sect:PRISM_via} turn to stochastic and variational approximations of PRISM and suggest algorithmic solutions.
This is followed by a set of numerical experiments in Section~\ref{sect:sim}, and then by  conclusions in Section~\ref{sect:con}.

\section{Preliminaries}
\label{sect:prelim}

\subsection{Notations}

Most of our notations are standard.
Vectors and matrices are represented by boldfaced lowercase and capital letters, e.g., $\bx$ and $\bX$, respectively (resp.);
unless otherwise specified, $\bx_i$ denotes the $i$th column of $\bX$;
$\Rbb, \Rbb_+, \Rbb_{++}$ are the sets of all real, non-negative and positive numbers, resp.;
given $\setX \subseteq \Rbb^n$, $\by \in \Rbb^n$, we denote $\setX + \by = \{ \bx + \by \mid \bx \in \setX \}$;
the superscripts $^\top$, $^{-1}$ and $^\dag$ denote transpose, inverse and pseudo-inverse, resp.;
$\bx = (x_1,\ldots,x_n)$ means that $\bx = [~ x_1, \ldots, x_n ~]^\top$;
$\| \cdot \|$ denotes the Euclidean norm;
$\tr( \cdot )$ is the trace of a matrix;
$\Diag(\bx)$ is a diagonal matrix whose $(i,i)$th element is $x_i$;
$\bzero$ is an all-zero vector;
$\bone$ is an all-one vector;
$\bI$ is an identity matrix;
$\be_i$ is a unit vector, i.e., $\be_i= (0,\ldots,0,1,0,\ldots,0)$ with $1$ being at the $i$th element;
given $\bA \in \Rbb^{m \times n}$,
\begin{align*}
	\sspan(\bA) & = \{ \by = \bA \bx \mid \bx \in \Rbb^n \}, \\
	\aff(\bA) & = \{ \by = \bA \bx \mid \bx \in \Rbb^n, \bone^\top \bx = 1 \}, \\
	\conv(\bA) & = \{ \by = \bA \bx \mid \bx \in \Rbb^n_+, \bone^\top \bx = 1 \}
\end{align*}
denote the span, affine hull and convex hull of $\{ \ba_1,\ldots, \ba_n \}$, resp.;
as a less standard notation,
\[
\bconv(\bA)  = \{ \by = \bA \bx \mid \bx \in \Rbb^n_{++}, \bone^\top \bx = 1 \}
\]
denotes the open convex hull of $\{ \ba_1,\ldots, \ba_n \}$ (it is open on $\aff(\bA)$, not on $\Rbb^m$);
$\indfn{\setX}$ is the indicator function of a set $\setX$:
\[
\indfn{\setX}(\bx) = \left\{
\begin{array}{ll} 1, & \bx \in \setX  \\ 0, & \bx \notin \setX \end{array}
\right. ;
\]
$\bx \sim D$ means that $\bx$ is a random variable with distribution $D$;
$\Exp_{\bx \sim p}[ \cdot ]$ denotes expectation of a random variable $\bx$ with distribution $p$;
$\var(x)$ and $\cov(\bx)$ denote the covariance of a random variable $x$ and $\bx$, resp.

The following specialized notations will be used frequently.
Given $\bA \in \Rbb^{m \times n}$, $\bs \in \Rbb^n$, we define, resp.,
\beq \label{eq:barA}
\bar{\bA} = [~ \ba_1 - \ba_n, \ldots, \ba_{n-1} - \ba_n ~], \quad
\bar{\bs} = (s_1,\ldots,s_{n-1}).
\eeq

\subsection{Simplex}

To describe simplex, we need to first review affine independence.
Let $\bA \in \Rbb^{m \times n}$.
We say that $\bA$ is affinely independent if $\bar{\bA}$ (cf. \eqref{eq:barA})
has full column rank.
A simplex is defined as a convex hull $\conv(\bA)$ with affinely independent $\bA$.
Additionally, a simplex is called full-dimensional if $m= n-1$.
A simplex $\conv(\bA)$ has the following properties:
its set of vertices is $\{ \ba_1,\ldots, \ba_n \}$;
its volume, according to \cite{gritzmann1995largestj}, is 
\beq \label{eq:svol}
\svol(\bA) = \frac{( \det( \bar{\bA}^\top \bar{\bA} ) )^{1/2}}{(n-1)!}.
\eeq

\subsection{Dirichlet Distribution}
\label{sec:Dir}

The Dirichlet distribution is commonly used to model on-unit-simplex random variables \cite{frigyik2010introduction,ng2011dirichlet}.
Let
\[
\Delta = \{ \bs \in \Rbb_+^N \mid \bone^\top \bs = 1 \},
\quad 
\bar{\Delta} = \{ \bs \in \Rbb_{++}^N \mid \bone^\top \bs = 1 \}
\]
be the unit simplex of $\Rbb^N$ and its open counterpart, resp.
A random variable $\bs \in \Delta$ is said to be Dirichlet distributed with concentration parameter $\balp \in \Rbb_{++}^N$, or simply $\balp$-Dirichlet distributed, if $\bar{\bs}= (s_1,\ldots,s_{N-1})$ 
has the density
\begin{align}
	D(\bar{\bs}; \balp) & =
	\frac{1}{B(\balp)}
	\left( \prod_{i=1}^{N-1} s_i^{\alpha_i - 1} \right) \left(1- \sum_{i=1}^{N-1} s_i \right)^{\alpha_N-1}   \indfn{\tilde{\Delta}}(\bar{\bs}),
	\label{eq:Dir_rig}
\end{align}
where 
\beq
\tilde{\Delta}  = \{ \bar{\bs} \in \Rbb_{++}^{N-1} \mid 1- \bone^\top \bar{\bs} > 0 \}; \nonumber 
\eeq
$B(\balp) = ( \prod_{i=1}^N \Gamma(\alpha_i) )/ \Gamma(\sum_{i=1}^N \alpha_i)$;
$\Gamma(x) = \int_0^\infty t^{x-1} e^{-t} {\rm d}t$  is the Gamma function.
Note that $\bar{\bs}$ is the truly operating random variable, as the last element $s_N = 1- \bone^\top \bar{\bs}$ of $\bs$ is completely determined by $\bar{\bs}$.
For convenience, however, it is common to write
\beq
D(\bs; \balp) = \frac{1}{B(\balp)} \left( \prod_{i=1}^{N} s_i^{\alpha_i - 1} \right) \indfn{\bar{\Delta}}(\bs)
\eeq
and write $\bs \sim D(\cdot; \balp)$ to specify a Dirichlet random variable.
The parameter $\balp$ governs the shape of the Dirichlet distribution, and the reader is referred to the literature \cite{frigyik2010introduction,ng2011dirichlet} for illustrations.
A well-known case is
\[
D(\bs; \bone ) =
(N-1)! \cdot
\indfn{\bar{\Delta}}(\bs),
\]
which is
the uniform unit-simplex distribution.

The Dirichlet distribution has a number of friendly properties.
First, it is easy to generate samples from it \cite{frigyik2010introduction}.
Second, many of its moments admit explicit expressions \cite{ng2011dirichlet}.
\begin{Fact}[{Dirichlet moments; see, e.g., \cite{ng2011dirichlet}}] \label{fac:Dir_moments}
	Let $\bs \sim D( \cdot; \balp)$. We have
	\begin{enumerate}[(a)]
		\item $\Exp[ \bs ] = \tilde{\balp}$, where $\tilde{\balp} = \balp/\alpha_0$, $\alpha_0 = \sum_{i=1}^N \alpha_i$;
		\item the covariance of $\bs$ is
		\[
		\cov(\bs) = \frac{1}{1+\alpha_0} ( \Diag(\tilde{\balp}) - \tilde{\balp} \tilde{\balp}^\top );
		\]
		\item the entropy of $\bs$ equals
		\begin{align*}
			H(\bs)  & := \Exp[ - \log D(\bs;\balp) ] \\
			&
			= \log B(\balp) - \sum_{i=1}^N (\alpha_i - 1 )( \psi(\alpha_i) - \psi(\alpha_0) ),
		\end{align*}
		where
		$\psi(x) = \frac{{\rm d} \log \Gamma(x)}{{\rm d} x}$ is the digamma function.
	\end{enumerate}
\end{Fact}
Third, the Dirichlet distribution can be used to construct distributions on a simplex.
\begin{Fact}[uniform distribution on a full-dimensional simplex] \label{fact:sim_dist}
	Let $\bx = \bB \bs$, where $\bB \in \Rbb^{(N-1) \times N}$ is affinely independent and $\bs \sim D(\cdot, \bone)$.
	The PDF of $\bx$ is
	\beq
	p(\bx) = \frac{1}{\svol(\bB)} \indfn{\bconv(\bB)}(\bx).
	\eeq
\end{Fact}
The proof of Fact \ref{fact:sim_dist} is shown in Appendix \ref{sect:proof fact 2}.

We will encounter integration involving the Dirichlet distribution.
The problem, in a general sense, is to integrate a function $f : \Rbb^N \rightarrow \Rbb$ over $\setA = \{ \bs \in \Rbb^N \mid \bone^\top \bs = 1 \}$.
A proper way to do so is
\beq \label{eq:clum_int}
\int_{\Rbb} \cdots \int_{\Rbb} f(s_1,\ldots,s_{N-1},{\textstyle 1- \sum_{i=1}^{N-1} s_i}) {\rm d} s_1 \cdots {\rm d} s_{N-1},
\eeq
where we apply $s_N = 1- \sum_{i=1}^{N-1} s_i$ directly.
Writing out \eqref{eq:clum_int} is clumsy, and we will use the Lebesgue integral
\[
\int f(\bs) {\rm d}\mu(\bs)
\]
to compactly represent~\eqref{eq:clum_int};
here $\mu$ is the Lebesgue measure on $\setA $.

\section{PRISM Formulation}
\label{sect:formulation}


As described in the Introduction and as illustrated in Fig.~\ref{fig:scatterplots},
we consider the following problem: 
We have a collection of data points $\by_1,\ldots,\by_T \in \Rbb^M$ that are posited to distribute on a simplex.
The vertices undergirding the simplex are unknown, and there is noise in the data points. 
Our aim is to identify the vertices from the data points.
In our probabilistic \SCA\ approach, or PRISM, we model the data points as
\beq \label{eq:base_model}
\by_t = \bx_t + \bv_t, \quad \bx_t = \bA_0 \bs_t \in \conv(\bA_0), \quad t = 1,\ldots,T,
\eeq
where 
$\bx_t \in \Rbb^M$ is the noise-free part of $\by_t$;
$\bA_0 \in \Rbb^{M \times N}$ 
collects the vertices of the simplex and is called the {\em true vertex matrix};
$\bs_t \in \Delta$ is a latent variable; 
$\bv_t \in \Rbb^M$ is noise.
The model \eqref{eq:base_model} is accompanied by the following assumptions:
\begin{Asm}
	The matrix $\bA_0$ is affinely independent.
\end{Asm}
\begin{Asm}
	The latent variables $\bs_t$'s are independently and identically distributed (i.i.d.).
	Every $\bs_t$ is uniformly distributed on $\bar{\Delta}$, or, $\bone$-Dirichlet distributed.
\end{Asm}
\begin{Asm}
	The noise variables $\bv_t$'s are i.i.d. and independent of the $\bs_t$'s.
	Every $\bv_t$ is Gaussian distributed with mean $\bzero$ and covariance $\sigma^2 \bI$, $\sigma > 0$.
\end{Asm}
Note that we consider a basic model wherein the $\bs_t$'s are treated as uniformly distributed random variables.
Such model is arguably reasonable when we have no prior information on the latent-variable distribution.



PRISM considers the following ML inference 
\beq \label{eq:ML}
\hat{\bA}_{\sf ML} \in  \arg
\max_{\bA \in \Rbb^{M \times N} } \mathcal{L}_T(\bA) : = \frac{1}{T} \sum_{t=1}^T \log p(\by_t; \bA),
\eeq
where $p(\by;\bA)$ is the PDF of a data point $\by$ parameterized by $\bA$.
Under the above data model, $p(\by;\bA)$ is given by
\begin{align}
	p(\by;\bA) & = \int p(\by|\bs; \bA) p(\bs) {\rm d}\mu(\bs) \nonumber \\
	& = (N-1)! \int \varphi_\sigma(\by - \bA\bs) \indfn{\bar{\Delta}}(\bs) {\rm d}\mu(\bs), \label{eq:pyA}
\end{align}
where $\varphi_\sigma(\by) = e^{- \| \by \|^2/{2\sigma^2} }/( \sqrt{2 \pi} \sigma)^{M}$ is a multivariate i.i.d. Gaussian function;
$p(\bs)$ is the PDF of a latent variable $\bs$;
$p(\by|\bs; \bA)$ is the PDF of a data point $\by$ conditioned on $\bs$ and parameterized by $\bA$.
There is no known closed-form solution for the integral \eqref{eq:pyA} in general.
The intractability of \eqref{eq:pyA} presents a challenge for realizing the ML estimator \eqref{eq:ML}, which we shall address.

The reader may wonder:
How about the ML alternative of maximizing the log likelihood over both $\bA$ and $\bs_1,\ldots,\bs_T$? To be precise, consider modeling $\bS = [~ \bs_1,\ldots,\bs_T ~]$ as a deterministic unknown with simplex support $\Delta^T = \{ \bS \in \Rbb^{N \times T} \mid \bs_t \in \Delta, \forall t \}$, and deal with the ML estimator
\[
\max_{\bA \in \Rbb^{M \times N}, \bS \in \Delta^T} \log p(\bY;\bS,\bA),
\]
where 
$p(\bY;\bS,\bA)$ is the PDF of $\bY= [~ \by_1,\ldots, \by_T ~]$ parameterized by $\bA$ and $\bS$.
It is easy to show that the above ML problem equals
\beq \label{eq:ML2}
\min_{\bA \in \Rbb^{M \times N}, \bS \in \Delta^T} \| \bY - \bA \bS \|^2,
\eeq
which 
is a simplex-structured matrix factorization (SSMF) problem and looks easier to handle than the ML estimator \eqref{eq:ML}.
But there is an issue.
\begin{Fact} \label{fac:ident_ML2}
	Let $\bR$ be any invertible matrix in $\Delta^N$. If $(\bA,\bS)$ is a solution to the SSMF problem \eqref{eq:ML2}, then $(\bA \bR^{-1}, \bR\bS)$ is also a solution to the SSMF problem \eqref{eq:ML2}.
\end{Fact}
The proof of Fact~\ref{fac:ident_ML2} is trivial and omitted for brevity.
Fact~\ref{fac:ident_ML2} indicates that, even without noise,
a solution to  problem \eqref{eq:ML2} does not necessarily equal the true vertex matrix $\bA_0$ or its column permuted counterparts.\footnote{
	Note that Fact~\ref{fac:ident_ML2} assumes general $\bA$.
	In the NMF context it is known that if $\bA$ is non-negative and we incorporate non-negative constraints with $\bA$ in the matrix factorization problem \eqref{eq:ML2}, then \eqref{eq:ML2} may provide some form of identifiability guarantees; see, e.g.,  \cite{fu2019nonnegative} and the references therein.}

\section{ML Identifiability}
\label{sect:id}

While the ML estimator \eqref{eq:ML2} fails to guarantee identifiable solutions with the vertices,
we will show that the more difficult ML estimator \eqref{eq:ML} can provide  identifiable solutions.

\subsection{The Identifiability Result}
The identifiability problem in question is classic in statistical inference.
Consider $T \rightarrow \infty$ such that, by the law of large numbers, the log likelihood function $\mathcal{L}_T$ in \eqref{eq:ML} converges to
\begin{align*}
	\mathcal{L}(\bA) & = \Exp_{\by \sim p(\cdot;\bA_0)}[ \log p(\by;\bA) ]
	= \int_{\Rbb^M} p(\by;\bA_0)  \log p(\by;\bA) {\rm d}\by.
\end{align*}
Consider the ML problem
\beq \label{eq:ML_bigT}
\max_{\bA \in \Rbb^{M \times N}} \mathcal{L}(\bA),
\eeq
which may intuitively be seen as the ML problem \eqref{eq:ML} for large data size $T$.
By Kullback-Leibler divergence, we have
\beq \label{eq:id_0}
\mathcal{L}(\bA_0) \geq \mathcal{L}(\bA),
\eeq
where equality in \eqref{eq:id_0}  holds if and only if
\beq \label{eq:id_1}
p(\by;\bA_0) =  p(\by;\bA), \quad \text{for all $\by$.}
\eeq
Eqs.~\eqref{eq:id_0}--\eqref{eq:id_1} suggest that the true vertex matrix $\bA_0$ is an ML solution in \eqref{eq:ML_bigT}, and $\bA$ is an ML solution in \eqref{eq:ML_bigT} if and only if \eqref{eq:id_1} holds.
Hence, our identifiability problem is to confirm whether \eqref{eq:id_1} does {\em not} hold for any non-trivial choice of $\bA$.
Our identifiability result is shown below.
\begin{Theorem} \label{thm:id}
	Eq.~\eqref{eq:id_1} holds if and only if $\bA = \bA_0 \bPi$, where $\bPi$ 
	is a permutation matrix.
	Consequently, $\bA$ is a solution to the ML problem \eqref{eq:ML_bigT} if and only if $\bA = \bA_0 \bPi$.
\end{Theorem}

Note that the underlying assumptions with Theorem~\ref{thm:id} are Assumptions 1--3 and $T \rightarrow \infty$.
Theorem~\ref{thm:id} confirms that the ML estimator \eqref{eq:ML_bigT} can exactly identify the vertices.
It also gives an intuitive implication that the finite-data ML estimator \eqref{eq:ML} should suppress the impact of noise better as we have more data points.
In the next subsection we will show an intuitive proof of Theorem~\ref{thm:id} to provide insight.
The formal proof of Theorem~\ref{thm:id} is relegated to Appendix \ref{sect:formal identifiability}.

\begin{Remark}
	It is natural to question whether the PRISM identifiability result in Theorem~\ref{thm:id} provides new insights compared to the known CG identifiability results \cite{arora2016computing,AGH12,recht2012factoring,gillis2014fast,ge2015intersecting,lin2018maximum,lin2015identifiability,Fu2015,fu2016robust}.
	Analyses in CG and PRISM operate under different assumptions (one deterministic, another stochastic), and it is hard to compare fairly.
	Still, let us compare one aspect, namely, whether we can reduce the impact of noise by increasing the data size $T$.
	The currently available analyses in CG 
	are unable to confirm the aforementioned aspect, although they can confirm how noise-robust an algorithm is in the worst-case sense and for any $T$.
	The PRISM identifiability result in Theorem~\ref{thm:id}, in comparison, requires infinite $T$ but can confirm elimination of the noise effects under infinite $T$.
\end{Remark}

%

\subsection{An Intuitive Proof of ML Identifiability}
\label{sect:proof_intuit}

The intuitive proof of Theorem~\ref{thm:id} is as follows.
From \eqref{eq:pyA}, it is easy to see that $\bA = \bA_0 \bPi$ implies $p(\by; \bA) = p(\by; \bA_0)$.
To show the converse,
consider the special case of  $M= N-1$ and affinely independent $\bA$.
By Fact~\ref{fact:sim_dist},
the noise-free components $\bx_t$'s in \eqref{eq:base_model} follow a uniform simplex distribution
\beq \label{eq:pxA}
p(\bx; \bA) = \frac{1}{\svol(\bA)} \indfn{\bconv(\bA)}(\bx).
\eeq
Applying \eqref{eq:pxA} to the model \eqref{eq:base_model}, we can write
\beq \label{eq:pyA_sim}
p(\by;\bA) = \int_{\Rbb^{N-1}} \varphi_\sigma(\by- \bx) p(\bx; \bA) {\rm d}\bx.
\eeq
By defining the Fourier transform (FT) of $f: \Rbb^n \rightarrow \Rbb$ as $\check{f}(\bxi)= \int_{\Rbb^n} f(\bx) e^{-j 2\pi \bxi^\top \bx} {\rm d}\bx$ and the inverse FT as $\int_{\Rbb^n} \check{f}(\bxi) e^{j 2\pi \bxi^\top \bx} {\rm d}\bxi$,
we have
\begingroup
\allowdisplaybreaks
\begin{subequations}
	\begin{align}
		& p(\by; \bA) = p(\by; \bA_0) \text{~ for all $\by$}
		\label{eq:inproof_0}
		\\
		\Longrightarrow ~  & \check{\varphi}_\sigma(\bxi) \check{p}(\bxi; \bA) = \check{\varphi}_\sigma(\bxi) \check{p}(\bxi; \bA_0) \text{~ for all $\bxi$}
		\label{eq:inproof_1} \\
		\Longrightarrow ~  & \check{p}(\bxi; \bA) = \check{p}(\bxi; \bA_0) \text{~ for all $\bxi$}
		\label{eq:inproof_2} \\
		\Longrightarrow ~  & p(\bx; \bA) = p(\bx; \bA_0) \text{~ for all $\bx$}
		\label{eq:inproof_3} \\
		\Longrightarrow ~  & \bconv(\bA) = \bconv(\bA_0)
		\label{eq:inproof_4} \\
		\Longrightarrow ~  & \conv(\bA) = \conv(\bA_0)
		\label{eq:inproof_5} \\
		\Longrightarrow ~  & \{ \ba_1,\ldots, \ba_N \} = \{ \ba_{0,1},\ldots,\ba_{0,N} \}.
		\label{eq:inproof_6}
	\end{align}
\end{subequations}
\endgroup
Here, \eqref{eq:inproof_1} is obtained by taking FT on both sides of  \eqref{eq:inproof_0} and by noting the convolution relation in \eqref{eq:pyA_sim};
\eqref{eq:inproof_2} is due to the fact that $\check{\varphi}_\sigma(\bxi)= e^{- 2 \pi^2 \| \bxi \|^2} \neq 0$ for all $\bxi$;
\eqref{eq:inproof_3} is obtained by taking inverse FT on both sides of \eqref{eq:inproof_2};
\eqref{eq:inproof_4} is the direct consequence of \eqref{eq:pxA};
\eqref{eq:inproof_5} is obtained by taking closure on both sides of  \eqref{eq:inproof_4};
\eqref{eq:inproof_6} is due to the fact that the set of all vertices of  $\conv(\bA)$ is $\{ \ba_1, \ldots, \ba_N \}$.
Our intuitive proof is complete.

The above intuitive proof takes insight from the ICA identifiability proof in \cite[Theorem~1]{khemakhem2020variational}, particularly, the FT and inverse FT steps in \eqref{eq:inproof_1}--\eqref{eq:inproof_3}.
In the formal proof, shown in Appendix \ref{sect:formal identifiability},
we will generalize the result to any $\bA$ and to any $M \geq N-1$.
Also we will fix a subtle issue---namely, using inverse FT to obtain \eqref{eq:inproof_3} is not rigorous.
As $p(\bx;\bA)$ is discontinuous, $\check{p}(\bxi; \bA)$ may not be integrable and its inverse FT may not exist.

\section{Relationships Between PRISM and SVMin}
\label{sect:relate}

Having pinned down the identifiability of PRISM, 
we continue by exploring the connection of PRISM and the SVMin approach in convex geometry.
Our study will focus on the case of $M=N-1$;
such assumption can be justified and has been used in the literature (e.g., \cite{Jose12,Ma2014HU}), and it will also be discussed in Appendix \ref{sect:app_DR}.

\subsection{SVMin is PRISM in the Noiseless Case}
\label{sect:connect_noisefree_SVMin}

In the preceding section,
we showed in \eqref{eq:pxA}--\eqref{eq:pyA_sim} that the PDF $p(\by;\bA)$ in the  case  of $M=N-1$ takes the form
\beq \label{eq:pyA_sim2}
p(\by;\bA) = \frac{1}{\svol(\bA)}
\int_{\Rbb^{N-1}} \varphi_\sigma(\by- \bx) \indfn{\bconv(\bA)}(\bx) {\rm d}\bx,
\eeq
for an  affinely independent $\bA$.
Let $\setA$ be the set of all $(N-1) \times N$ affinely independent matrices,
and restrict the ML problem \eqref{eq:ML} as
\beq \label{eq:ML3}
\max_{\bA \in \setA} \mathcal{L}_T(\bA) = \frac{1}{T} \sum_{t=1}^T \log p(\by_t; \bA)
\eeq
so that \eqref{eq:pyA_sim2} applies.
Consider the noiseless case where
\[
\log p(\by; \bA) = - \log \svol(\bA) + \log( \indfn{\bconv(\bA)}(\by) ).
\]
Since
\[
\log( \indfn{\bconv(\bA)}(\by) ) = \left\{ 
\begin{array}{ll}
	0, & \by \in \bconv(\bA) \\
	-\infty, & \by \notin \bconv(\bA)
\end{array} 
\right. 
\]
we may rewrite \eqref{eq:ML3} as 
\beq
\begin{aligned}
	\min_{\bA \in \setA} & \; \log \svol(\bA) \\
	{\rm s.t.}  & \; \by_t \in \bconv(\bA),
	\quad t=1,\ldots,T.
\end{aligned} \label{eq:MVES_from_ML}
\eeq 
We see that problem \eqref{eq:MVES_from_ML} aims to find a data enclosing simplex $\conv(\bA)$ that yields the minimum volume---which is SVMin \cite{Chan2009}.
We thereby have
the revelation that {\em PRISM reduces to SVMin in the noiseless case.}

The above identity was mentioned by
Bioucas-Dias in his WHISPERS 2009 oral presentation \cite{nascimento2009learning}.
It was not shown in his papers \cite{nascimento2009learning,nascimento2012hyperspectral}, although, for experts, it was alluded to.
A somewhat similar result was also mentioned by Dobigeon {\em et al.} \cite[Appendix]{dobigeon2009joint}.

\subsection{Connection to Volume-Regularized Matrix Factorization}
\label{sect:connect_SVMin_SSMF}

\begin{figure}[h]
	\centering
	\begin{subfigure}[b]{0.33\linewidth}
		\includegraphics[width=\textwidth]{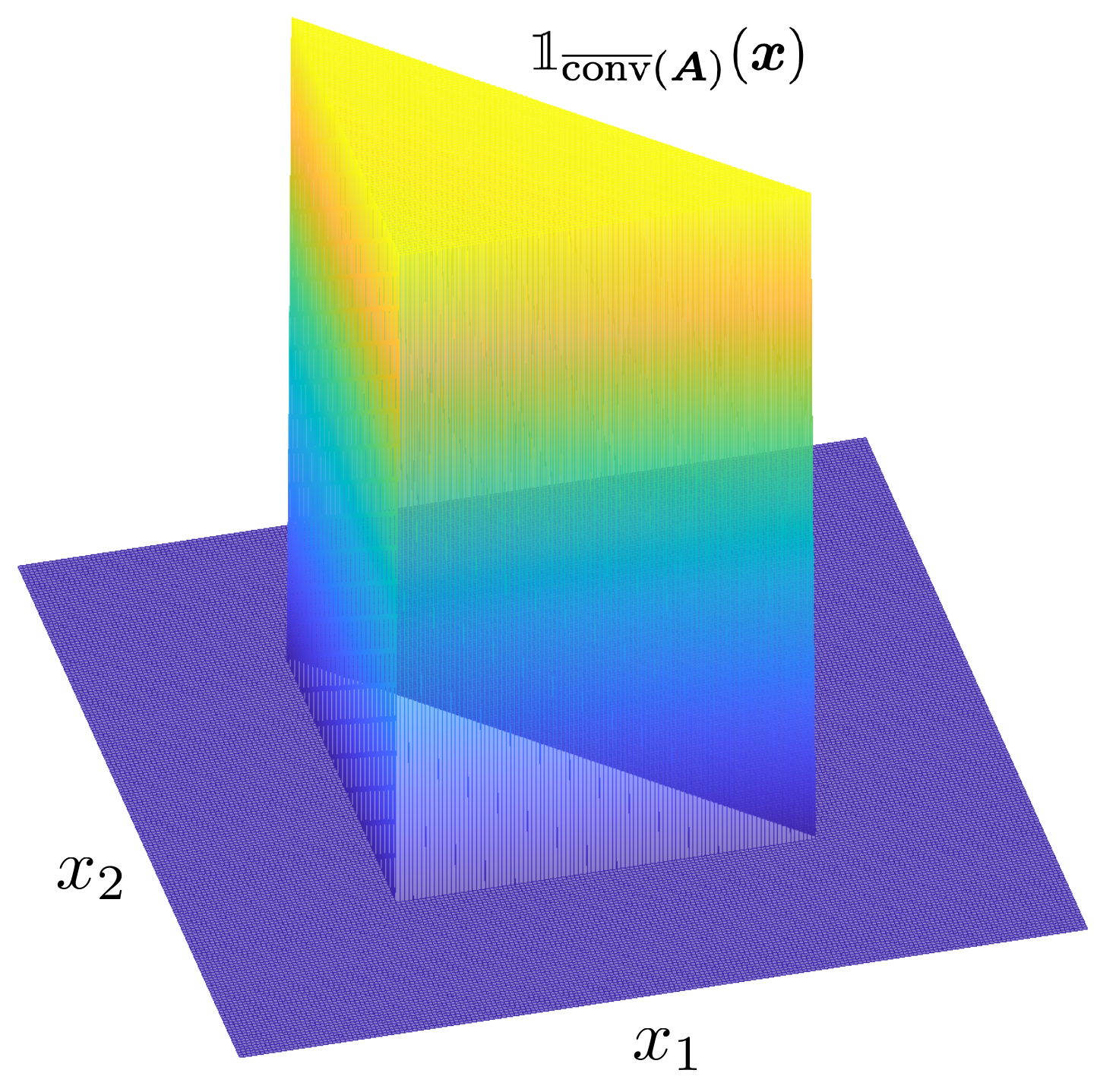}
		\caption{}
	\end{subfigure}
	\hfil
	\begin{subfigure}[b]{0.33\linewidth}
		\includegraphics[width=\textwidth]{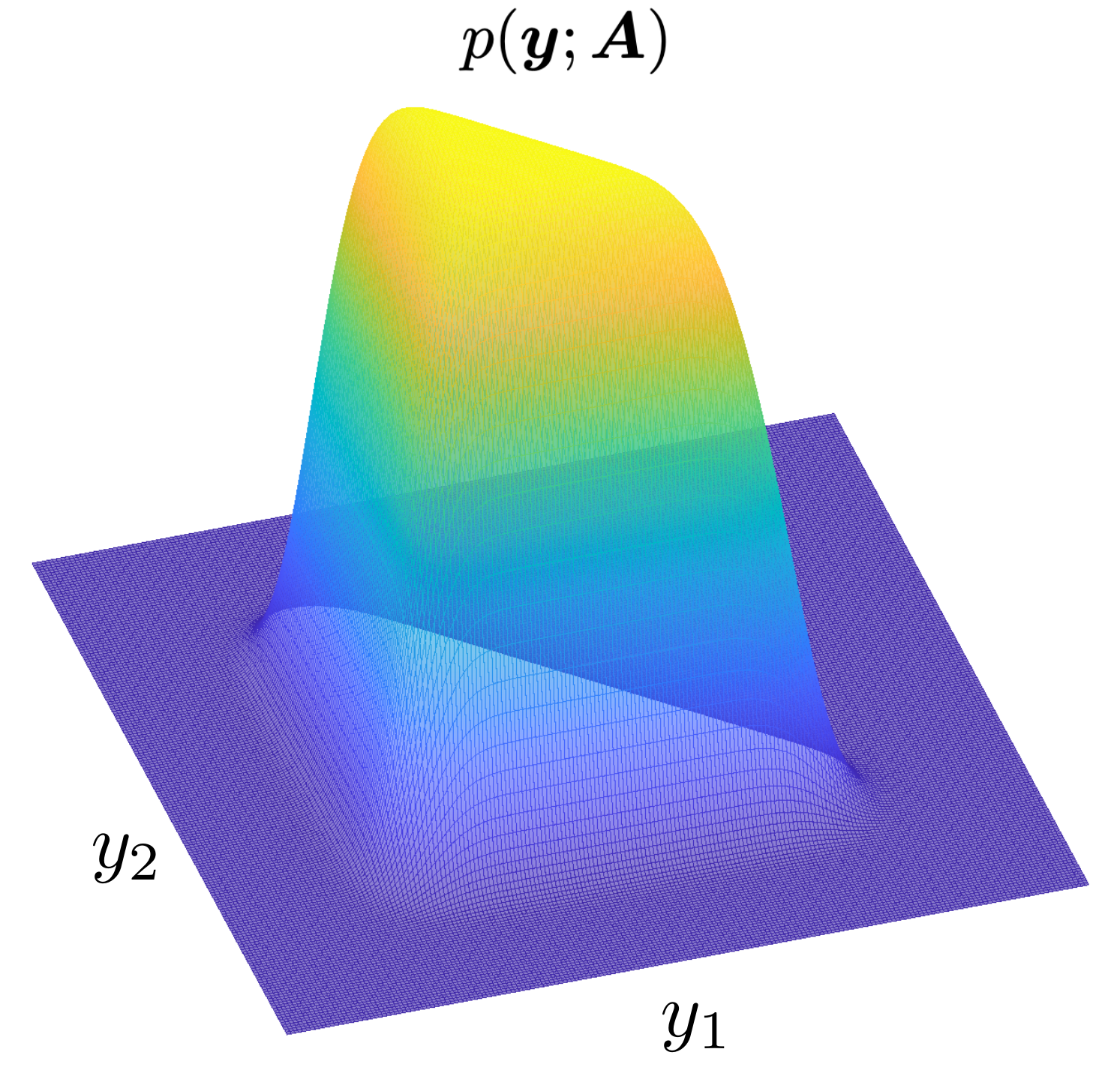}
		\caption{}
	\end{subfigure}
	\caption{Illustration of $\indfn{\bconv(\bA)}(\bx)$ and $p(\by;\bA)$ in \eqref{eq:pyA_sim2}.}
	\label{fig:PDF}
\end{figure}

Let us turn to the noisy case.
From \eqref{eq:pyA_sim2}, 
we see the following:
$p(\by;\bA)$ is the convolution of a multivariate Gaussian function $\varphi_\sigma(\by)$ and the on-off function $\indfn{\bconv(\bA)}(\by)$ 
(ignoring the scale $1/\svol(\bA)$);
see Fig.~\ref{fig:PDF}.
Or, 
$p(\by;\bA)$ is a ``blurred'' version of $\indfn{\bconv(\bA)}(\by)$ with
edges being smoothed.
This observation leads us to 
consider 
an edge-smooth approximation
\beq \label{eq:e_smooth}
p(\by; \bA) \approx \frac{1}{\svol(\bA)} e^{-  \frac{1}{\lambda} \cdot {\rm dist}(\by, \conv(\bA))^2},
\eeq
where ${\rm dist}(\bx,\setX) := \inf_{\bx' \in \setX} \| \bx - \bx' \|$ is the distance of a point $\bx$ and a set $\setX$;
$\lambda > 0$ determines the smoothness level which should scale with $\sigma^2$.
The right-hand side of \eqref{eq:e_smooth} mimics $p(\by;\bA)$ in the sense that it is constant if $\by \in \conv(\bA)$, and it gradually goes down as $\by$ moves away from $\conv(\bA)$.
We argue that \eqref{eq:e_smooth} is reasonable for high SNRs.

Let us examine the approximation of the ML problem~\eqref{eq:ML3} under \eqref{eq:e_smooth}: 
\begin{align}
	- \max_{\bA \in \setA} \mathcal{L}_T(\bA) 
	  & 
	\approx
	\min_{\bA \in \setA}  \log \svol(\bA) +  \frac{1}{ \lambda T} \sum_{t=1}^T {\rm dist}(\by_t, \conv(\bA))^2
	\nonumber  \\
	& 
	=   \min_{\bA \in \setA, \bS \in {\Delta}^T} \log \svol(\bA) +  \frac{1}{ \lambda T} \| \bY - \bA \bS \|^2;
	\label{eq:VRMF}
\end{align}
note ${\rm dist}(\by,\conv(\bA))^2= \min_{\bs \in {\Delta}} \| \by - \bA \bs \|^2$.
Problem \eqref{eq:VRMF} appears as a volume-regularized SSMF,
a typical altered formulation of the 
noiseless SVMin formulation
\eqref{eq:MVES_from_ML} in the noisy case \cite{miao2007endmember,fu2016robust}.
For convenience, we will call \eqref{eq:VRMF} {\em SVMin-SSMF.}
To conclude, we can interpret SVMin-SSMF as an approximate PRISM.
Intuitively, the approximation should be good for high SNRs.

\subsection{Connection to Soft-Constrained SVMin}
\label{sect:connect_SISAL}

By the same argument as above, 
we can also see SISAL  \cite{Dias2009}, a popularly-used algorithm in SVMin, as an approximate PRISM.
To put into context, note the following result.
\begin{Fact}[{polyhedral form of a full-dimensional simplex \cite{Chan2009}}] \label{fac:sim2poly}
	Let $\bA \in \Rbb^{(N-1) \times N}$ be affinely independent.
	We have
	\[
	\conv(\bA) = \textstyle \cap_{i=1}^N \setH_i(\bA),
	\]
	where each $\setH_i(\bA)  = \{ \by \mid \bb_i^\top \by \geq c_i \}$
	is a halfspace;
	\begin{align}
		\bB & := [~ \bb_1,\ldots\bb_{N-1}~] =  \bar{\bA}^{-\top}, 
		\label{eq:change_of_var1} \\
		\bc & := (c_1,\ldots,c_{N-1}) =  \bar{\bA}^{-1} \ba_N,
		\label{eq:change_of_var2} \\
		c_N & :=  - \bone^\top \bc - 1, ~ \bb_N := - \bB\bone. \label{eq:change_of_var_cond}
	\end{align}
\end{Fact}
Our intuition is to build a variant of the approximation \eqref{eq:e_smooth} by penalizing points that lie outside $\setH_i(\bA)$.
To this end, consider a change of variables
$\bA \in \setA \rightarrow (\bB,\bc) \in \setB \times \Rbb^{N-1}$ according to \eqref{eq:change_of_var1}--\eqref{eq:change_of_var2},
where $\setB$ denotes the set of all invertible $(N-1) \times (N-1)$ matrices.
We adopt the following approximation 
\beq \label{eq:e_smooth_sisal}
p(\by;\bA) \approx \frac{1}{\svol(\bA)} e^{-  \frac{1}{\lambda} \textstyle \sum_{i=1}^N ( c_i - \bb_i^\top \by )_+ }  
\eeq 
for some $\lambda > 0$,
where $( x )_+ := \max\{ 0, x \}$;
$( c_i - \bb_i^\top \by )_+$ serves as a penalty function for the violation of $\by \in \setH_i(\bA)$.
The resulting approximation of the ML problem \eqref{eq:ML3} is 
\beq \label{eq:SISAL}
\begin{aligned} 
	\min_{\bB, \bc} & ~  - \log |\det(\bB)| +  \frac{1}{\lambda T} \sum_{t=1}^T \sum_{i=1}^N
	( c_i - \bb_i^\top \by_t )_+ \\
	{\rm s.t.} & ~ \bB \in \setB, ~ \text{\eqref{eq:change_of_var_cond} holds}.
\end{aligned}
\eeq
Problem \eqref{eq:SISAL} is similar to the formulation of SISAL \cite{Dias2009},
whose rationale is to replace the hard constraints $\by \in \conv(\bA)$ in the original SVMin formulation \eqref{eq:MVES_from_ML}  with ``soft constraints'' 
to make the solution robust against noise.

\subsection{Connection to Chance-Constrained SVMin}
\label{sect:connect_chance_SVMin}

We can also draw a connection to robust SVMin via chance constraints \cite{Arul2011}. 
Observe
\begin{align*}
	\int_{\Rbb^{N-1}} \varphi_\sigma(\by- \bx) \indfn{\conv(\bA)}(\bx) {\rm d}\bx
	& = {\rm Prob}( \by - \bv \in \conv(\bA) ),
\end{align*}
where $\bv \sim \varphi_\sigma$.
Using the polyhedral representation in Fact~\ref{fac:sim2poly}, we get
\begin{align*}
	{\rm Prob}( \by - \bv \in \conv(\bA) )
	& \leq {\rm Prob}( \by - \bv \in \setH_i(\bA) ) \\
	& = {\rm Prob}( -\bb_i^\top \bv \geq c_i - \bb_i^\top \by  ) \\
	& =  \Phi\left( \frac{\bb_i^\top \by - c_i }{\sigma \| \bb_i \| } \right)
\end{align*}
for any $i$, where $\Phi(x) = \frac{1}{\sqrt{2\pi}} \int_{-\infty}^x e^{-z^2/2} {\rm d}z$.
The above equations lead to 
an
upper-bound approximation
\beq \label{eq:upper_bnd_pyA}
p(\by;\bA) \leq \frac{1}{\svol(\bA)} \min_{i=1,\ldots,N}   \Phi\left( 
\frac{\bb_i^\top \by - c_i }{ \sigma \| \bb_i \| }
\right).
\eeq
By applying \eqref{eq:upper_bnd_pyA} and the change of variables in \eqref{eq:change_of_var1}--\eqref{eq:change_of_var2},
we obtain an approximation of the ML problem \eqref{eq:ML3} as follows
\beq \label{eq:aml_prob}
\begin{aligned} 
	\min_{\bB, \bc} & ~  - \log |\det(\bB)| + \frac{1}{T} \sum_{t=1}^T \max_{i=1,\ldots,N} - \log  \Phi\left(
	\frac{\bb_i^\top \by_t - c_i }{ \sigma \| \bb_i \|}
	\right) \\
	{\rm s.t.} & ~ \bB \in \setB, ~ \text{\eqref{eq:change_of_var_cond} holds}.
\end{aligned}
\eeq
Problem \eqref{eq:aml_prob}
is reminiscent of an existing SVMin formulation, namely, the chance-constrained SVMin
\beq \label{eq:mves_chance}
\begin{aligned}
	\min_{\bB \in \setB, \bc} & ~ - \log |\det(\bB)|  \\
	{\rm s.t.} & ~  \Phi( ( \bb_i^\top \by_t - c_i)/(\sigma \| \bb_i \|)) \geq \eta, ~ \text{for all $i,t$}
\end{aligned}
\eeq
for a pre-specified $\eta \in [0,1]$ \cite{Arul2011}. 
Problem \eqref{eq:mves_chance} was proposed as an alteration of the noiseless SVMin formulation \eqref{eq:MVES_from_ML},
wherein the data enclosing constraints of \eqref{eq:MVES_from_ML} are modified as chance constraints to improve robustness to noise. 
We see that by changing the penalty terms of problem \eqref{eq:aml_prob} as constraints, we
get the chance-constrained SVMin problem \eqref{eq:mves_chance}.

\subsection{Further Discussion}

In Appendix	\ref{sect:discuss_svmin}, we further discuss aspects arising from the PRISM-SVMin relationships revealed above.


\section{Importance Sampling Approximation (ISA)}
\label{sect:PRISM_isa}

Having shown the relationships of PRISM and SVMin in the previous section, we now turn our attention to designated schemes for realizing PRISM.
This section will consider ISA, while the next section  will be devoted to variational inference approximation (VIA).
Note that, unlike the previous section, we do not assume $M= N-1$.

\subsection{A Variational Reformulation of ML}

The ISA scheme to be presented is an instance of Monte Carlo expectation maximization (MCEM).
To describe, we consider a slightly non-standard presentation that will cover expectation maximization (EM), MCEM, and later, VIA.
Let $q$ be any PDF that is measurable on $\{ \bs \in \Rbb^N \mid \bone^\top \bs = 1 \}$ and has support $\bar{\Delta}$.
Let $p(\by,\bs;\bA) = p(\by|\bs;\bA) p(\bs)$.
Consider the Jensen inequality
\beq \label{eq:jen}
\begin{aligned}
	\log p(\by;\bA) & = \log \left( \int p(\by,\bs;\bA) \frac{q(\bs)}{q(\bs)} {\rm d}\mu(\bs)  \right)   \\
	& = \log\left(  \Exp_{\bs \sim q}[ p(\by,\bs;\bA)/q(\bs) ] \right)  \\
	& \geq \Exp_{\bs \sim q}[ \log( p(\by,\bs;\bA)/q(\bs) ) ] \\
	& := \hat{\ell}(\bA,q;\by),
\end{aligned}
\eeq
where equality in \eqref{eq:jen} holds if and only if
\beq \label{eq:trun_gaus}
q(\bs) \propto p(\by,\bs;\bA) \propto \varphi_\sigma(\by-\bA \bs) \indfn{\bar{\Delta}}(\bs),
\eeq
i.e., $q$ is a unit-simplex truncated Gaussian distribution.
Note that \eqref{eq:trun_gaus} is equivalent to
\beq \label{eq:q_jen}
q(\bs) = \frac{ p(\by,\bs;\bA) }{ \int p(\by,\bs;\bA) {\rm d}\mu(\bs) } = p(\bs|\by;\bA).
\eeq
Using \eqref{eq:jen}, we can reformulate the ML problem \eqref{eq:ML} as 
\beq \label{eq:ML_var}
\max_{\substack{ \bA \in \Rbb^{M \times N},  q_t \in \mathcal{D} \, \forall t }} \textstyle \hat{\mathcal{L}}_T(\bA,\{ q_t \}):= \frac{1}{T} \sum_{t=1}^T \hat{\ell}(\bA,q_t;\by_t),
\eeq
where $\setD$ is the family of all distributions with support $\bar{\Delta}$.

We should justify why we are interested in \eqref{eq:ML_var}, a seemingly more complex ML formulation.
The difficulty with the original ML problem \eqref{eq:ML} is that $\log p(\by;\bA)$ is an intractable integral.
We want to see if \eqref{eq:ML_var}, or its approximations, will circumvent the difficulty.
To put into context, consider an alternating maximization (AM) method for problem \eqref{eq:ML_var}:
\begin{subequations} \label{eq:am_base}
	\begin{align}
		\bA^{k+1}  & \in \arg \max_{\bA \in \Rbb^{M \times N}} \hat{\mathcal{L}}_T(\bA,\{ q_t^k \}), \label{eq:am_base_a} \\
		q_t^{k+1} & \in \arg \max_{q_t \in \setD } \hat{\ell}(\bA^{k+1},q_t;\by_t),
		\quad t=1,\ldots,T, \label{eq:am_base_b}
	\end{align}
\end{subequations}
for $k=0,1,\cdots$.
The ISA and VIA schemes to be developed seek two different ways to approximate \eqref{eq:am_base}.

\subsection{The ISA-PRISM Scheme}

Before we describe our ISA scheme, we should first note that the AM \eqref{eq:am_base} for realizing the ML is identical to EM.
By the Jensen inequality result in \eqref{eq:jen}--\eqref{eq:q_jen}, 
the solution to \eqref{eq:am_base_b} is $q_t^{k+1} = p( \cdot | \by_t; \bA^{k+1})$.
By putting this solution to \eqref{eq:am_base_a}, we can simplify the AM \eqref{eq:am_base} to
\beq \label{eq:em}
\bA^{k+1} \in \arg \max_{\bA \in \Rbb^{M \times N}} \textstyle \sum_{t=1}^T \Exp_{  \bs_t \sim p(\cdot | \by_t; \bA^k)}[ \log p(\by_t | \bs_t; \bA ) ].
\eeq 
Eq.~\eqref{eq:em} takes the same form as EM, which was derived by a minorization-maximization methodology.
In considering \eqref{eq:em}, we wish that either \eqref{eq:em} would be easy to solve, or \eqref{eq:em} would admit a tractable objective function.
But none of the above is true in our problem.
A natural idea is then to apply ISA, using a large amount of randomly drawn samples to approximate the objective function.
Such idea is identical to MCEM \cite{wei1990monte}.

Let us go into the details.
Let $\bxi_t^1,\ldots, \bxi_t^{R_t}$ be an $R_t$ number of randomly drawn samples from $p( \cdot | \by_t; \bA^{k})$.\footnote{Actually $\bxi_t^1,\ldots, \bxi_t^{R_t}$ and $R_t$ depend on the iteration $k$, but we shall suppress the latter for brevity.}
This requires us to generate samples from a unit-simplex truncated Gaussian distribution,
and it can be done by rejection sampling or MCMC methods; see, e.g., \cite{altmann2014sampling,cong2017fast}.
We apply the ISA
\[
\Exp_{ \bs_t \sim p(\cdot | \by_t; \bA^k)}[ \log p(\by_t | \bs_t; \bA ) ]
\approx 
\textstyle \frac{1}{R_t} \sum_{r=1}^{R_t} \log p(\by_t | \bxi_t^r; \bA ).
\]
The subsequent approximation of \eqref{eq:em} can be shown to be
\beq
\bA^{k+1} \in \arg \max_{\bA \in \Rbb^{M \times N}} - \textstyle \sum_{t=1}^T \frac{1}{R_t}  \sum_{r=1}^{R_t}  \| \by_t - \bA \bxi_t^r \|^2,
\eeq 
which is a least squares problem with solution
\begin{align}
	\bA^{k+1} & =
	\textstyle
	\left( \sum_{t=1}^T \by_t \bm m_t^\top  \right)
	\left( \sum_{t=1}^T\bR_t \right)^\dag, \label{eq:mcem} \\
	\bm m_t & =
	\textstyle
	\frac{1}{R_t} \sum_{r=1}^{R_t}  \bxi_t^{r},
	\quad  \bR_t  = \frac{1}{R_t} \sum_{r=1}^{R_t}  \bxi_t^{r} (\bxi_t^{r} )^\top. \nonumber
\end{align}
To summarize, our ISA-PRISM scheme is given by \eqref{eq:mcem}, where, at each iteration $k$, we generate $\{ \bxi_t^r \}$ from $p( \cdot | \by_t; \bA^{k})$ by a sampling method.

\subsection{Discussion}

Let us discuss the advantages and drawbacks of the above ISA-PRISM scheme.
The implementation of ISA-PRISM is very simple.
It can also deal with more complex models such as models under the presence of outlying data points, Dirichlet mixture models, nonlinear and variability models, non-negative $\bA$, etc.;
such extensions will not be pursued here, and the reader is referred to \cite{wu2017stochastic,wu2019stochastic,li2020stochastic}.
However, using ISA also means that we need a massive amount of samples to approximate well.
This is particularly a concern for large problem dimension $N$.
This issue is coupled with another issue, namely, the low efficiencies of known sampling methods for  large problem dimension $N$;
e.g., rejection sampling tends to reject many samples, or take many iterations to generate one sample, when $N$ is large.
Empirically we found that ISA-PRISM works very well for $N$ less than $10$, but performs poorly for larger $N$.
We should note that the aforementioned merits and limitations are common in Monte~Carlo-based inference methods.

\section{Variational Inference Approximation (VIA)}
\label{sect:PRISM_via}

The ISA-PRISM scheme in the last section implements the ML estimator by applying Monte Carlo approximation to the intractable integral in the alternating maximization  \eqref{eq:am_base}.
VIA attacks \eqref{eq:am_base} by a different route, namely, restricting the distribution family $\setD$ so that the objective function is tractable.

\subsection{Dirichlet VIA}

Consider restricting $\setD$, the family of all $\bar{\Delta}$-supported distributions, in the ML formulation \eqref{eq:ML_var} by the Dirichlet family
\beq \label{eq:D_choice}
\setD= \{ q = D(\cdot;\balp) \mid \balp \in \Rbb_{++}^N \}.
\eeq 
Such restriction will lead to a lower-bound approximation of the ML.
We have two reasons for this.
First, among all $\bar{\Delta}$-supported distributions, the Dirichlet distribution is the most well-understood.
Second, it can be shown that, under $q = D( \cdot; \balp )$, 
the function $\hat{\ell}(\bA,q;\by)$ in \eqref{eq:jen} can be written as
\begin{align}
	-\hat{\ell}(\bA,q;\by)  
	&
	\propto
	\frac{1}{2\sigma^2} \Exp[ \| \by - \bA \bs \|^2 ] - H(\bs) \nonumber  \\
	&  = \frac{1}{2\sigma^2} ( \| \by - \bA \Exp[ \bs ] \|^2 + \TVar(\bA \bs) ) - H(\bs)
	:= f(\bA, \balp; \by),
	\label{eq:ELBO_2}
\end{align}
where we denote $\Exp_{\bs \sim q}[ \cdot ] = \Exp[ \cdot ]$ for brevity;
$H(\bs) = \Exp[- \log q(\bs) ]$ is the entropy;
$\TVar(\bx) = \sum_{i=1}^n \var(x_i) =  \tr( \cov(\bx) )$.
We know from Fact~\ref{fac:Dir_moments} that $\Exp[\bs]$, $\cov(\bs)$ and $H(\bs)$ have  explicit expressions;
we will examine the details later.
Consequently, the ML problem \eqref{eq:ML_var} under the restrictive approximation \eqref{eq:D_choice}, or VIA-ML for short, has a tractable objective function.

\subsection{VIA-ML is Regularized Matrix Factorization}

It is worthwhile to pause a moment to draw connections.
From \eqref{eq:ML_var} and \eqref{eq:ELBO_2}, we can write the VIA-ML problem as 
\beq \label{eq:SSMF_VIA}
\min_{ \bTheta \in \setC }
\frac{1}{2 \sigma^2} ( \| \bY -  \bA \Exp[ \bS ] \|^2 + \TVar(\bA\bS)) - H(\bS),
\eeq
where $\bs_t \sim D(\cdot; \balp_t )$;
$\bTheta= \{ \bA, \balp_1,\ldots,\balp_T \}$;
$\setC = \Rbb^{M \times N} \times \Rbb^N_{++} \times \cdots \times \Rbb^N_{++}$;
$\TVar(\bX) = \sum_{j=1}^n \TVar(\bx_j)$;
$H(\bS) = \sum_{t=1}^T H(\bs_t)$.
We see that {\em the VIA-ML problem \eqref{eq:SSMF_VIA} resembles a matrix factorization problem}---the first term of \eqref{eq:SSMF_VIA} is a data fitting term in matrix factorizaton; the second term is a penalty term on variances, encouraging smaller variances; the third term is a negative-entropy penalty term,
discouraging smaller variances (entropy tends to be larger for more diversely distributed distributions).
Note that, from a matrix factorization viewpoint, problem \eqref{eq:SSMF_VIA} has no regularization parameter to tune.
We also show the following result:

\begin{Prop} \label{prop:SSMF2}
	If we remove the entropy term $H$ from the VIA-ML problem \eqref{eq:SSMF_VIA}, the resulting problem is equivalent to the plain SSMF
	\beq \label{eq:prob:SSMF20}
	\min_{\bA \in \Rbb^{M \times N}, \bXi \in \bar{\Delta}^T} \| \bY - \bA \bXi \|^2.
	\eeq
\end{Prop}
{\em Proof of Proposition~\ref{prop:SSMF2}:} \
Define $\eta_t = \bone^\top \balp_t$, $\bxi_t = \balp_t / \eta_t \in \bar{\Delta}$.
From Fact~\ref{fac:Dir_moments} one readily gets $\Exp[ \bs_t ]= \bxi_t$,
\begin{align*}
	\cov(\bs_t) & = \frac{1}{1+ \eta_t} \bC(\bxi_t), \quad
	\bC(\bxi_t)  = \Diag(\bxi_t) - \bxi_t \bxi_t^\top.
\end{align*}
Problem \eqref{eq:SSMF_VIA} without $H$ can be expressed as
\beq \label{eq:prob:SSMF2}
\min_{\bTheta \in \setC}
f(\bTheta) := \| \bY - \bA \bXi \|^2 + \textstyle \sum_{t=1}^T \frac{1}{1+ \eta_t} \tr(\bA \bC(\bxi_t) \bA^\top ).
\eeq
Let
$\bTheta^*$
be a solution to \eqref{eq:prob:SSMF2}.
Let $\eta_t^* = \bone^\top \balp_t^*$, $\bxi_t^* = \balp_t^*/ \eta_t^*$,
and choose $\balp_t = \eta_t \bxi_t^*$ for any $\eta_t > \eta_t^*$.
Then we see from \eqref{eq:prob:SSMF2} that $\bTheta = \{ \bA^*,\balp_1, \ldots, \balp_T\}$ has $f(\bTheta^*) \geq f(\bTheta)$.
This implies that there always exists a solution for which $\eta_t \rightarrow \infty$ for all $t$.
As the second term of $f$ in \eqref{eq:prob:SSMF2} diminishes as $\eta_t \rightarrow \infty$,
we can reduce \eqref{eq:prob:SSMF2} to \eqref{eq:prob:SSMF20}.
This completes the proof.
\hfill $\blacksquare$

\begin{Remark}
	Proposition \ref{prop:SSMF2} reveals a limitation.
	Suppose $\sigma^2$ is very small.
	The entropy term $H$ in the VIA-ML problem \eqref{eq:SSMF_VIA} would have negligible effects, and, by Proposition \ref{prop:SSMF2}, the VIA-ML problem should be close to the plain SSMF.
	Moreover the plain SSMF is an unidentifiable formulation, as indicated  
	in Fact~\ref{fac:ident_ML2}.
	This implies that VIA-ML may not work well for high SNRs.
	Our empirical results to be presented seem to be in agreement with the above argument.
	But our empirical results will also indicate that VIA-ML works well for low SNRs.
\end{Remark}

\subsection{The VIA-PRISM Scheme}

We now turn to algorithmic realization.
Our VIA-PRISM scheme is the realization of AM \eqref{eq:am_base} under the aforementioned VIA.
By applying Fact~\ref{fac:Dir_moments},
we express the right-hand side of \eqref{eq:ELBO_2}  as
\begingroup
\allowdisplaybreaks
\begin{subequations} \label{eq:f_def}
	\begin{align}
		f(\bA,\balp;\by) & = g(\bA,\balp,\bone^\top \balp;\by) + \textstyle \sum_{i=1}^N h(\alpha_i) + \iota(\bone^\top \balp) + C,
		\\
		g(\bA,\balp,\eta;\by) & =
		\tfrac{1}{\sigma^2} \left( - \tfrac{1}{\eta} \by^\top \bA\balp + \tfrac{1}{2 (1+\eta) \eta} \tr(\bA \bR(\balp) \bA^\top) \right), \label{eq:fn_g} \\
		\bR(\balp) & = \Diag(\balp) + \balp \balp^\top,
		\\
		h(\alpha) & = -\log \Gamma(\alpha) + (\alpha-1) \psi(\alpha), \label{eq:fn_h} \\
		\iota(\eta) & =  \log \Gamma(\eta) - (\eta-N) \psi(\eta),
	\end{align}
\end{subequations}
\endgroup
where $C$ is a constant; 
recall that 
$\Gamma(x) = \int_0^\infty t^{x-1} e^{-t} {\rm d}t$ is the Gamma function, and  $\psi(x) = \frac{{\rm d} \log \Gamma(x)}{{\rm d} x}$ is the digamma function.
The AM \eqref{eq:am_base} under the VIA can be written as
\begin{subequations}
	\begin{align}
		\bA^{k+1} & \in \arg \min_{\bA \in \Rbb^{M \times N}} \textstyle \sum_{t=1}^T g(\bA,\balp_t^k,\bone^\top \balp_t^k;\by_t),  \label{eq:via_am_a} \\
		\balp_t^{k+1} & \in \arg \min_{\balp \in \Rbb^N_{++}} f(\bA^{k+1},\balp;\by_t),
		\quad t=1,\ldots,T. \label{eq:via_am_b}
	\end{align}
\end{subequations}
Problem~\eqref{eq:via_am_a} is a least squares problem with solution
\beq 
\label{eq:via_am_a_sol}
\bA^{k+1} = \left[ \sum_{t=1}^T \tfrac{1}{\eta_t^k} \by_t (\balp_t^k)^\top \right] \left[ \sum_{t=1}^T \tfrac{1}{(1+ \eta_t^k) \eta_t^k} \bR(\balp_t^k) \right]^{-1},
\eeq
where $\eta_t^k = \bone^\top \balp_t^k$.
The problems in \eqref{eq:via_am_b} are not easy and will be treated next.

\subsection{Are the Variational Problems in \eqref{eq:via_am_b} Solvable?}

For notational convenience, let us rewrite \eqref{eq:via_am_b} as
\beq \label{eq:VIA_prob}
\min_{\balp \in \Rbb_{++}^N} f(\bA,\balp;\by),
\eeq
Problem \eqref{eq:VIA_prob} is non-convex; the term $\eta= \bone^\top \balp$ in \eqref{eq:fn_g} is particularly troublesome.
As a fundamental study, we beg this question:
can problem \eqref{eq:VIA_prob} be solvable?
To answer that, we reformulate problem \eqref{eq:VIA_prob} as
\beq \label{eq:VIA_prob_eq}
\min_{\eta \in \Rbb_{++}} r(\eta) + \iota(\eta),
\eeq
where
\beq \label{eq:cvx_VIA_sub}
\begin{aligned}
	r(\eta)= \min_{\balp \in \Rbb_{++}^N} & \, g(\bA,\balp,\eta;\by) + \textstyle \sum_{i=1}^N h(\alpha_i) \\
	{\rm s.t.} & \, \bone^\top \balp = \eta.
\end{aligned}
\eeq
{\em Suppose} that $r(\eta)$ is efficiently computable for any given $\eta > 0$.
Then we may argue that problem \eqref{eq:VIA_prob_eq} is not that difficult---for 
we can use grid search to find the solution to problem \eqref{eq:VIA_prob_eq} (up to an accuracy).
In practice it is more pragmatic to employ line search, rather than grid search.
We would expect, at least intuitively, that the chance for line search to find the optimal solution to a one-dimensional problem should be high.
Hence the question boils down to whether problem \eqref{eq:cvx_VIA_sub} is efficiently solvable.
Observe that $g$ in \eqref{eq:fn_g} is convex in $\balp$.
If $h$ is convex on $\Rbb_{++}$ then  problem \eqref{eq:cvx_VIA_sub} is convex.
We show that this is true.
\begin{Prop} \label{prop:cvx_VIA_sub}
	The function $h$ is strictly convex on $\Rbb_{++}$. As a direct corollary, problem \eqref{eq:cvx_VIA_sub} is strictly convex.
\end{Prop}
The proof of Proposition~\ref{prop:cvx_VIA_sub} is relegated to Appendix \ref{sect:app_cvx_VIA}.
We also custom-develop an efficient solver for problem \eqref{eq:cvx_VIA_sub} via the  augmented direction method of multipliers (ADMM) \cite{boyd2011distributed}; the details are also relegated to Appendix \ref{sect:app_cvx_VIA_admm}.

Let us summarize how we numerically solve problem \eqref{eq:VIA_prob}.
There are two levels.
The first level applies line search to problem \eqref{eq:VIA_prob_eq}; we employ Golden search.
At each line search iteration, the computation of $r(\eta)$ for a specific $\eta$ is required.
This is done at the second level, where we solve problem \eqref{eq:cvx_VIA_sub} by the ADMM solver in  Appendix \ref{sect:app_cvx_VIA_admm}.
The pseudo code is provided in Algorithm~\ref{alg:VIA-PRISM}.

%
%
%

\begin{algorithm}[htb!]
	\caption{Solver for the varational problem in \eqref{eq:VIA_prob}} \label{alg:VIA-PRISM}
	\begin{algorithmic}[1]
		\State \textbf{given}: a search interval $ [a,b] $
		\State set $ [\eta_1,\eta_2]=[a,b] $, $ \mu=\frac{1+\sqrt{5}}{2} $ (the golden ratio)
		\Repeat
		\State  let $ \eta_3 = \eta_2-(\eta_2-\eta_1)/\mu $, $ \eta_4 = \eta_1+(\eta_2-\eta_1)/\mu $
		\State  compute $ r(\eta_3) $ and $ r(\eta_4) $ by solving problem \eqref{eq:cvx_VIA_sub} using the ADMM solver in Appendix \ref{sect:app_cvx_VIA_admm}
		\If {$ r(\eta_3)+\iota(\eta_3) < r(\eta_4)+\iota(\eta_4) $} 
		\State update $ \eta_2=\eta_4 $ 
		\Else  
		\State{update $ \eta_1=\eta_3 $ }
		\EndIf
		
		\Until a stopping rule is satisfied
		\State set $ \eta = (\eta_1+\eta_2)/2 $
		\State find the solution $\balp^\star$ to problem \eqref{eq:cvx_VIA_sub} using the ADMM solver in Appendix \ref{sect:app_cvx_VIA_admm}
		\State {\bf output}  $\balp^\star$
	\end{algorithmic}
\end{algorithm}

\section{Numerical Experiments}
\label{sect:sim}

We performed numerical experiments to examine the potential of PRISM.

\begin{table*}[ht]
	\centering
	\resizebox{\linewidth}{!}{%
	\begin{tabular}{c|c|c|c}
		\hline\hline
		algorithm & formulation \& reference            & initialization           & parameter settings \\ \hline\hline
		SVMAX     & pure-pixel search  \cite{Chan2011}                     & /                        & /                   \\ \hline
		SISAL     & \multirow{3}{*}{soft-constrained SVMin, cf. \eqref{eq:MVES_from_ML} \cite{Dias2009}}   &
		\multirow{3}{*}{VCA \cite{Nascimento2005}}     
		& volume regularization parameter $ \lambda = 0.1 $      \\ \cline{1-1} \cline{4-4} 
		SISAL-t   &  		  &  			 & 
		\begin{tabular}[l]{@{}l@{}}MSE vs SNR: $\lambda=0.02\times({\rm SNR}-7)/3$ ($ N=5 $) \\
			MSE vs SNR: $\lambda=0.02\times{\rm SNR}/3$ ($ N=20 $)\\ 
			MSE vs $ T $: $\lambda=\frac{0.01}{[T/1000]+1}$ (SNR=10dB)\\
			MSE vs $ T $: $\lambda=\frac{0.2}{[T/1000]+1}$ (SNR=20dB)\\
		\end{tabular} \\ \hline
		RVolMin   & SVMin-SSMF in \eqref{eq:rvolmin} \cite{fu2016robust}     & SISAL-t                  & 
		$\lambda=0.5,p=0.5$
		\\ \hline
		MVES      & noiseless SVMin in \eqref{eq:MVES_from_ML}  \cite{Chan2009}        & solve a feasibility problem & /                   \\ \hline
		RMVES     & chance-constrained SVMin in \eqref{eq:mves_chance} \cite{Arul2011} & VCA             & $\eta=0.16 $        \\ \hline
		ISA-PRISM & ML, importance sampling approx., Section~\ref{sect:PRISM_isa}                    & SVMAX                   & rejection sampling from $500$ samples             \\ \hline
		VIA-PRISM & ML, variational inference approx., Section~\ref{sect:PRISM_via}                  & SVMAX                   & /                   \\ \hline
	\end{tabular}}
	\caption{Settings of the various schemes.}
	\label{tab:algo_settings}
\end{table*}

\subsection{Algorithm Settings}

The algorithm settings of the ISA-PRISM and VIA-PRISM schemes  in Sections \ref{sect:PRISM_isa} and \ref{sect:PRISM_via} are described as follows.
Both ISA-PRISM and VIA-PRISM deal with non-convex optimization, and a reasonable initialization for them would be desirable. 
We initialize the two schemes by
SVMAX~\cite{Chan2011}, a computationally light CG algorithm by pure-pixel search.\footnote{SVMAX belongs to a representative type of pure-pixel search methods. It resembles VCA  \cite{Nascimento2005}, a very widely-used pure-pixel search algorithm.
	It is nearly identical to SPA \cite{gillis2014fast}, which is equipped with noise robustness analyses.}
ISA-PRISM is implemented by rejection sampling.
Specifically, for each data point $\by_t$, we give $R=500$ random samples and pick up the accepted samples by rejection sampling.
Note that the number of accepted samples $R_t$ for each data point $\by_t$ can vary from one point to another.
For VIA-PRISM, we set the parameter $\rho$ of the ADMM solver (see Appendix \ref{sect:app_cvx_VIA_admm}) as $0.01$.
We stop the ADMM solver when the dual error is less than $ 0.005$.
Furthermore, we stop ISA-PRISM and VIA-PRISM when their iteration numbers exceed $100$.

We benchmark PRISM against the following state-of-the-art schemes:
i) the pure-pixel search algorithm SVMAX;
ii) MVES~\cite{Chan2009}, which realizes the noiseless SVMin in \eqref{eq:MVES_from_ML};
iii) the famous SISAL~\cite{Dias2009}, which adopts a soft-constrained SVMin formulation similar to  \eqref{eq:SISAL};
iv) RVolMin~\cite{fu2016robust}, which considers the SVMin-SSMF formulation 
\beq \label{eq:rvolmin}
\min_{\bA, \bS \in {\Delta}^T} \textstyle  \lambda \log \det(\bA^\top \bA) +  \sum_{t=1}^T \| \by_t - \bA \bs_t \|^p
\eeq 
for some given $p > 0$, $\lambda > 0$;
v) RMVES~\cite{Arul2011}, which is based on the chance-constrained SVMin  formulation in \eqref{eq:mves_chance}.
Some key settings of the above schemes are shown in Table~\ref{tab:algo_settings}.
For SISAL we consider two implementations.
One, simply called ``SISAL'', has the regularization parameter $\lambda$ fixed for all experiments.\footnote{Here, $\lambda$ refers to the volume regularization parameter presented in the SISAL paper \cite{Dias2009}, not the one in the closely-related formulation \eqref{eq:SISAL}.}
Another, called ``SISAL-t'', has $\lambda$ manually tuned for better performance;  our tuning is heuristic and heavily empirical based.
Also,
we stop SISAL and RVolMin when the iteration numbers exceed $250$ and $1,000$, resp.;
we stop MVES and RMVES when the relative objective value changes are less than $ 10^{-8} $ and $ 10^{-6} $, resp.

\subsection{Synthetic Data Experiments}

We performed a collection of synthetic data experiments.
The data model \eqref{eq:base_model} and the accompanied assumptions are used to generate the data points.
In each simulation trial, we randomly generate $\bA_0$ by the element-wise independent $[0,1]$-uniform distribution.
We measure the estimation performance by mean square error (MSE)
${\sf MSE}(\bA_0, \hat{\bA})=\min_{\bm\pi\in\Pi_N}\frac{1}{MN}\sum_{n=1}^N\|{\ba}_{0,\pi_n}- \hat{\ba}_n\|^2,$
where $\hat{\bA}$ is the estimated vertex matrix; $\Pi_N$ is the set of all permutations of $\{1,2,\ldots,N\}$.
We fix $M= 50$, and we use $500$ independent trials to obtain the results below.

\subsubsection{Varying the Data Length $T$}
\begin{figure}[!th]
	\centering
	\begin{subfigure}[b]{0.32\linewidth}
		\includegraphics[width=\textwidth]{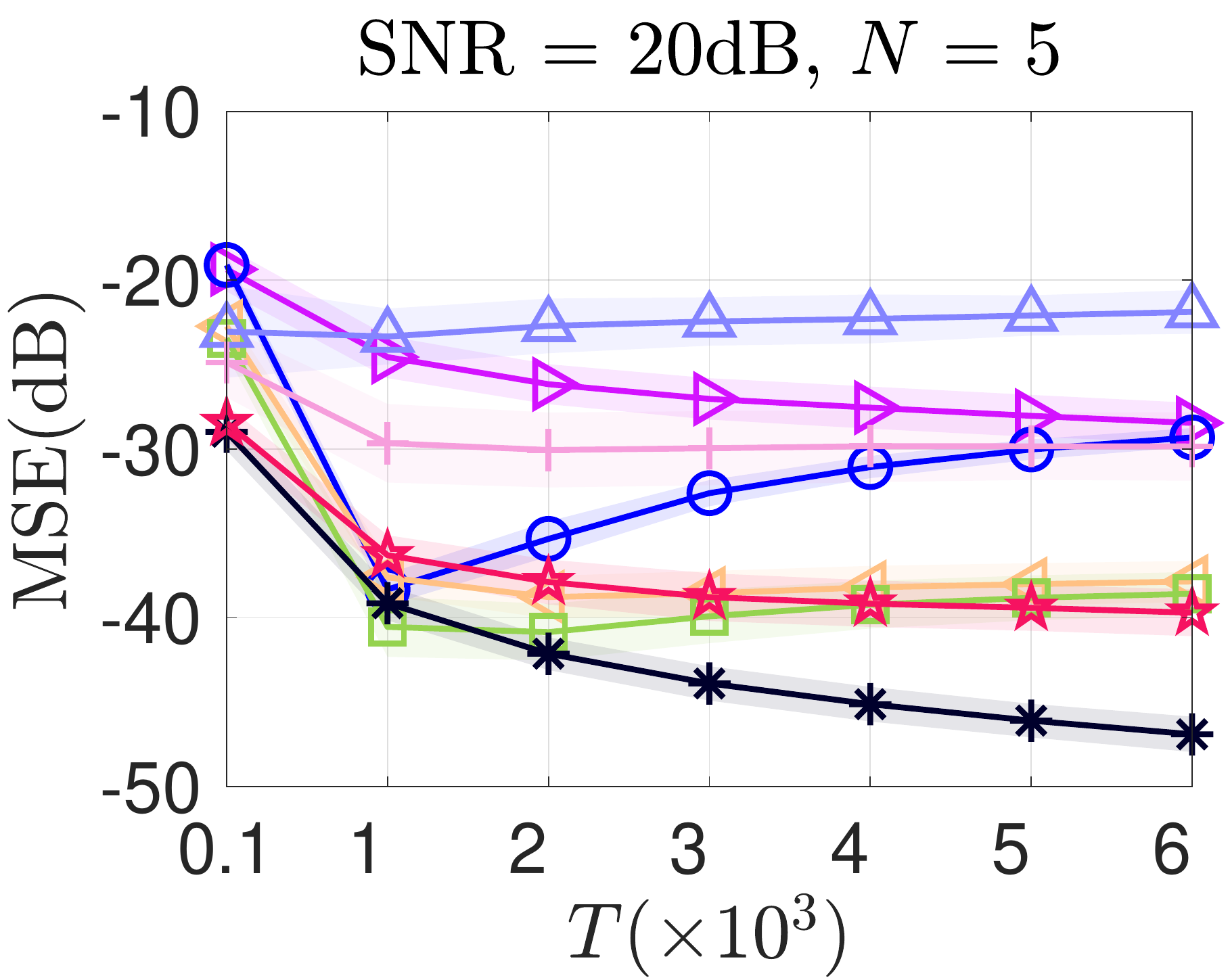}
	\end{subfigure}
	\begin{subfigure}[b]{0.32\linewidth}
		\includegraphics[width=\textwidth]{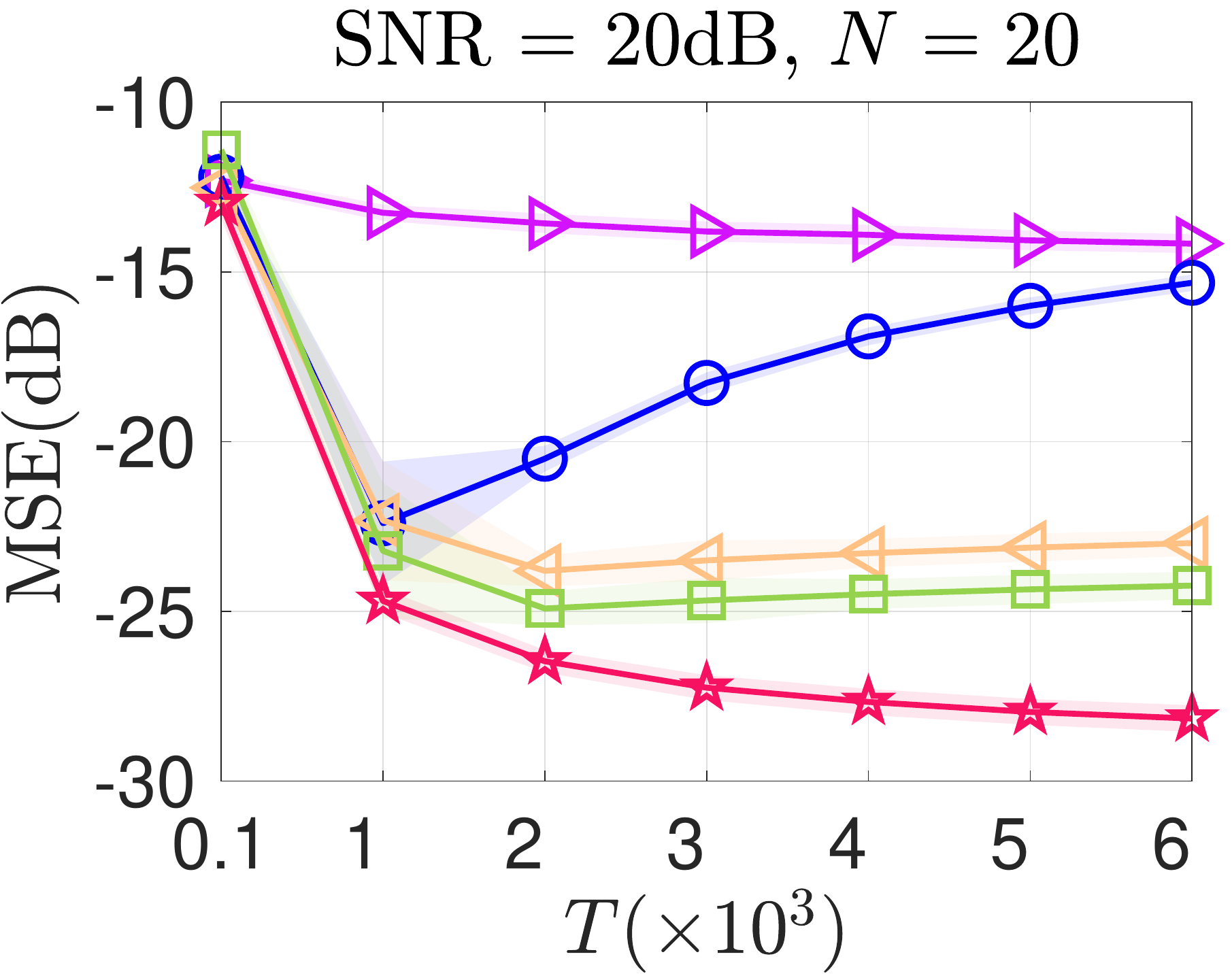}
	\end{subfigure}
	\begin{subfigure}[b]{0.32\linewidth}
		\includegraphics[width=\textwidth]{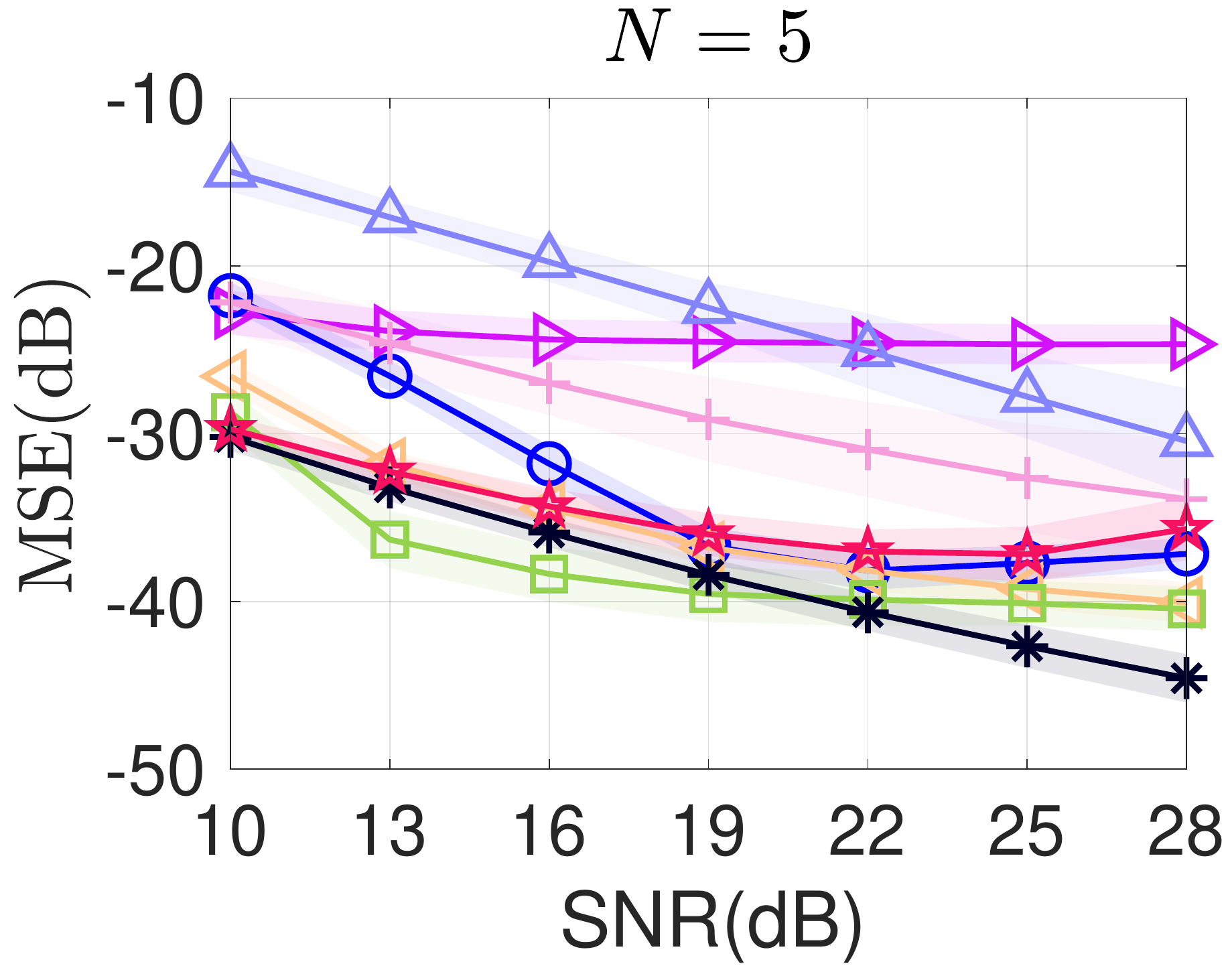}
	\end{subfigure}\vspace{0.6em}
	\begin{subfigure}[b]{0.32\linewidth}
		\includegraphics[width=\textwidth]{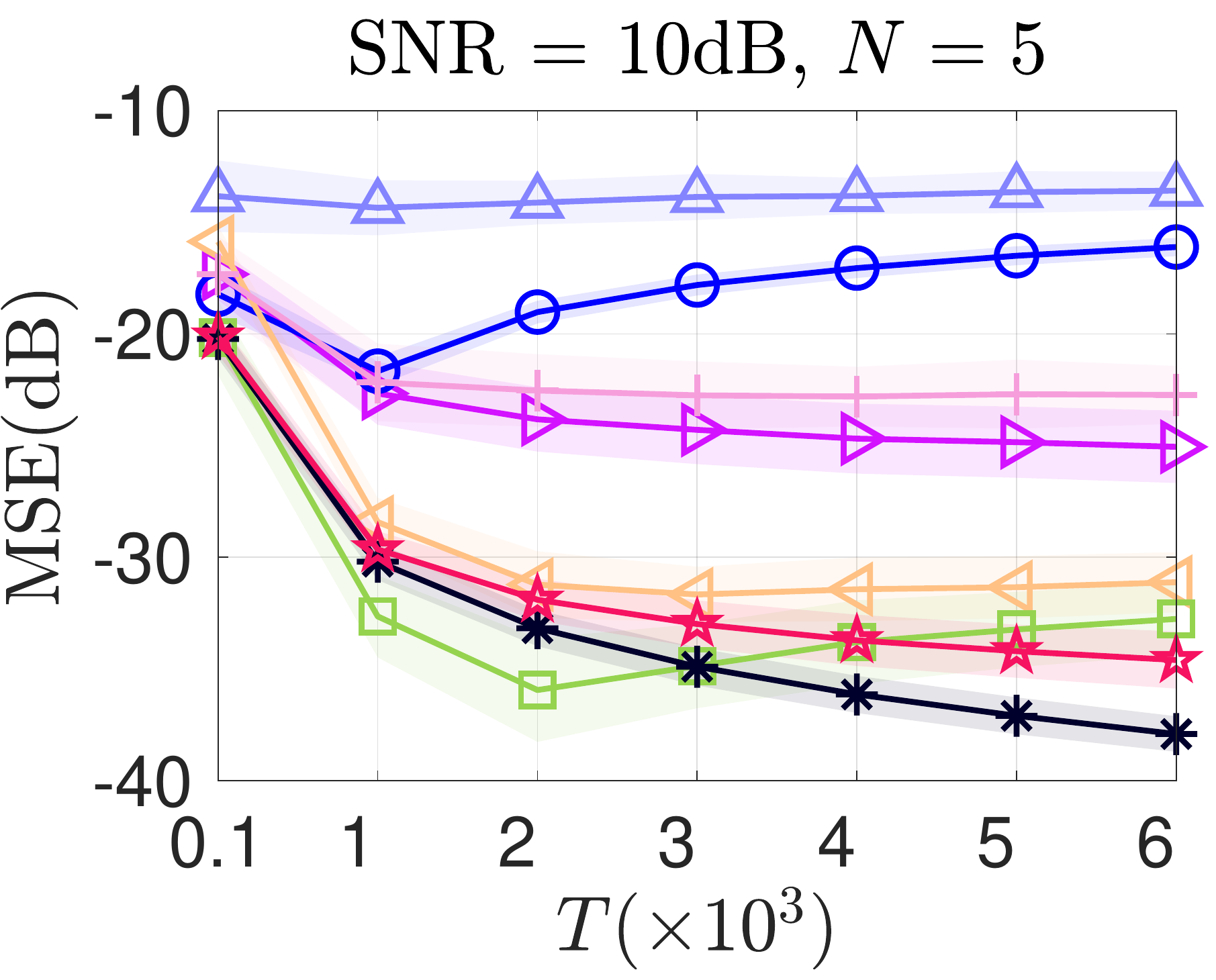}
		\caption{MSE versus $ T $ ($ N=5 $)}
		\label{fig:syn T N5}
	\end{subfigure}
	\begin{subfigure}[b]{0.32\linewidth}
		\includegraphics[width=\textwidth]{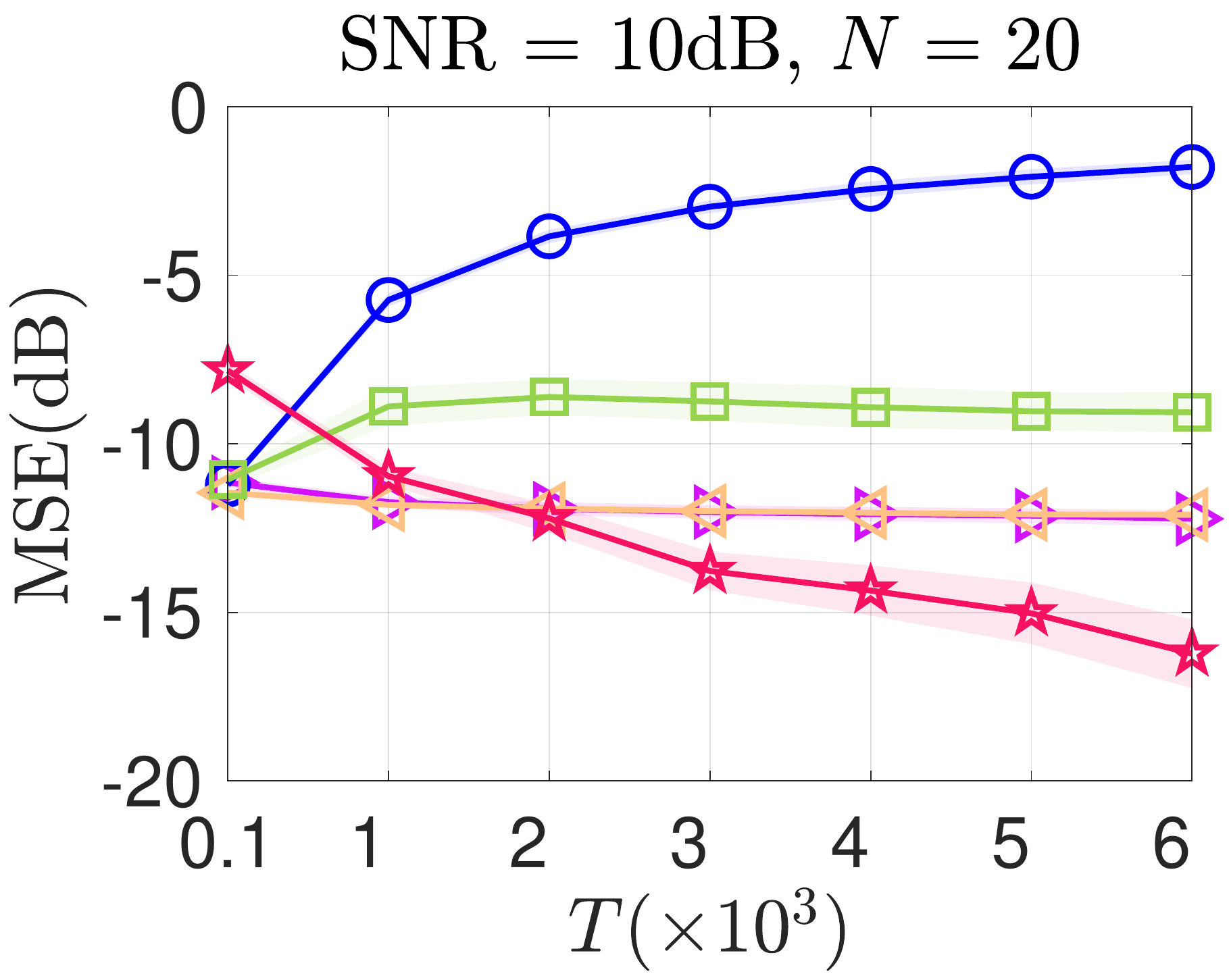}
		\caption{MSE versus $ T $ ($ N=20 $)}
		\label{fig:syn T N20}
	\end{subfigure}
	\begin{subfigure}[b]{0.32\linewidth}
		\includegraphics[width=\textwidth]{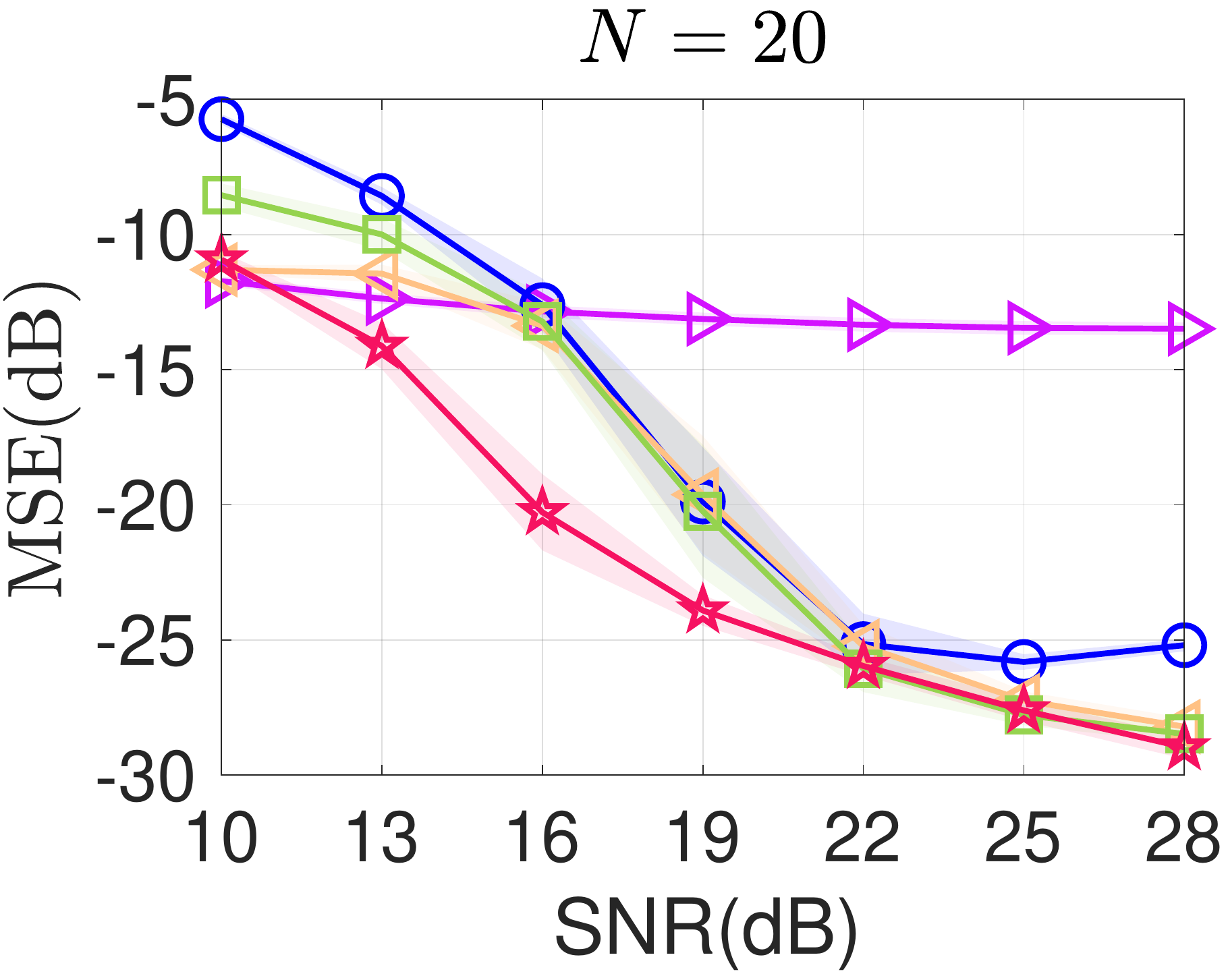}
		\caption{MSE versus SNR ($T=1,000$)}
		\label{fig:syn snr}
	\end{subfigure}\vspace{0.2em}
	\begin{subfigure}[b]{0.82\linewidth}
		\includegraphics[width=\textwidth]{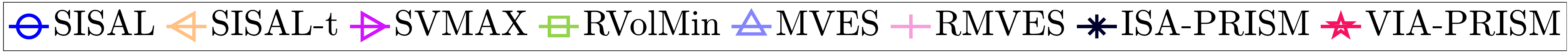}
	\end{subfigure}
	\caption{MSE performance. $M=50$.}
\end{figure}

We are interested in examining how the various schemes perform as the number of available data points, $T$, increases;
intuitively, one would expect that the MSE improves with $T$ in a consistent manner.
In Figs.~\ref{fig:syn T N5} and \ref{fig:syn T N20} we show the MSEs for different $T$.
Note that the lines are average MSEs, while the shadows indicate the standard deviations of the MSEs.
Let us first set our eyes on the case of $N= 5$ in Fig.~\ref{fig:syn T N5}.
We observe the following:
\begin{enumerate}[1.]
	\item First, the MSEs of ISA-PRISM and VIA-PRISM improve as $T$ increases.
	This is in agreement with the identifiability theorem in Theorem~\ref{thm:id}, which says that PRISM can perfectly identify the vertices when $T$ approaches infinity.
	That being said, we also notice that the MSE improvement is slow as $T$ becomes very large, say, $T \geq 4,000$.
	
	\item Second, the benchmark schemes do not seem to show consistent MSE improvement with $T$.
	As such, ISA-PRISM and VIA-PRISM are able to perform better than the benchmark schemes for large $T$.
	We also observe that SISAL, which has no parameter tuning, behaves peculiarly.
	SISAL-t, which has parameter tuning, gives better performance.
	The heuristic nature of the tuning should however be noted.
	
	\item Third, there is a performance gap between ISA-PRISM and VIA-PRISM.
	The reason could be with the approximation errors of VIA.
	The gap is seen to be larger for the higher SNR case (SNR= $20$dB).
	In this regard we should recall that, as indicated in Proposition~\ref{prop:SSMF2} and discussed in Remark 2, VIA may not work well for very high SNRs.
	
\end{enumerate}

Next, we turn to the case of $N= 20$ in Fig.~\ref{fig:syn T N20}.
We did not try MVES and RMVES because they run too slowly for large $N$.
We also did not consider ISA-PRISM because it fails to work; 
rejection sampling generates almost no sample.
We observe similar performance behaviors as the case of $N=5$.
Moreover, VIA-PRISM is seen to perform better than the benchmark schemes for  large $T$.

%
%
%
%

\subsubsection{Varying the Noise Level}
Fig.~\ref{fig:syn snr} shows the MSEs for different SNRs.
Note that we fix $T=1,000$.
We see that, for $N= 5$, ISA-PRISM generally outperforms the other schemes.
For $N= 20$, VIA-PRISM generally works better than the benchmark schemes;
this is particularly so for lower SNRs.
In addition, it is worth mentioning that
SISAL-t gives reasonably good performance. 
But note that SISAL-t requires heuristic parameter tuning, while
VIA-PRISM does not.

%
%

%

\subsubsection{Runtime Comparison}
We compare the runtimes of some of the schemes in Table~\ref{tab:runtime}. 
We consider $T=1,000$, SNR= $10$dB,
and the runtimes were recorded on a desktop computer with Intel Core i7 3.20GHz CPU and 64GB memory, and under MATLAB2019.
The computational costs of ISA-PRISM and VIA-PRISM are seen to be on the high side.
We should say that this study focuses on fundamental aspects and sets aside efficient implementations. 
The latter will be future work.

\begin{table}[hbt]
	\caption{Average runtimes (sec.) of the various schemes.}
	\label{tab:runtime}
	\centering
		\begin{tabular}{|c||c|c|c|c|c|}
			\hline
			& SVMAX & SISAL   & RVolMin & ISA-PRISM & VIA-PRISM \\ \hline \hline 
			$N=5$ & 0.003 & 0.06    & 0.65  & 30.76    & 15.93 \\ \hline 
			$N=20$ & 0.005 & 0.27  & 1.37 & -         & 189.89 \\ \hline
		\end{tabular}
\end{table}

\subsection{Real Data Experiment}

We apply PRISM to real data.
The application of interest is hyperspectral unmixing (HU) in remote sensing.
The problem is well-known in the remote sensing literature (see, e.g., \cite{Jose12,Ma2014HU}),
and it is concisely described as follows.
We are given a hyperspectral image of a captured scene, which has a few hundreds of spectral bands and has high spectral resolution.
That image is represented by $\by_1,\ldots, \by_T$,
where each $\by_t \in \Rbb^M$ is a pixel collecting measurements over a number of $M$ spectral bands, and $T$ is the number of pixels. 
We posit that the $\by_t$'s follow the model in \eqref{eq:base_model},
where the columns of $\bA_0$ are the spectral signatures of different materials that underlie the scene, and $\bs_t$ describes the materials' distribution at pixel $t$.
The HU problem is to recover $\bA_0$ from the $\by_t$'s, thereby identifying the materials.

The dataset we use is the Cuprite AVIRIS dataset, taken in 1997 in the Cuprite area by airborne visible/infrared imaging spectrometer (AVIRIS)~\cite{vane1993airborne}.
It is widely used in HU.
Previous studies revealed that, for the Cuprite AVIRIS dataset, pure-pixel search already gives very good results \cite{Nascimento2005,Chan2011}.
Still, it is interesting to use this dataset to demonstrate whether PRISM can provide reasonable results.

\begin{figure}[h]
	\centering
	\includegraphics[width=.8\linewidth]{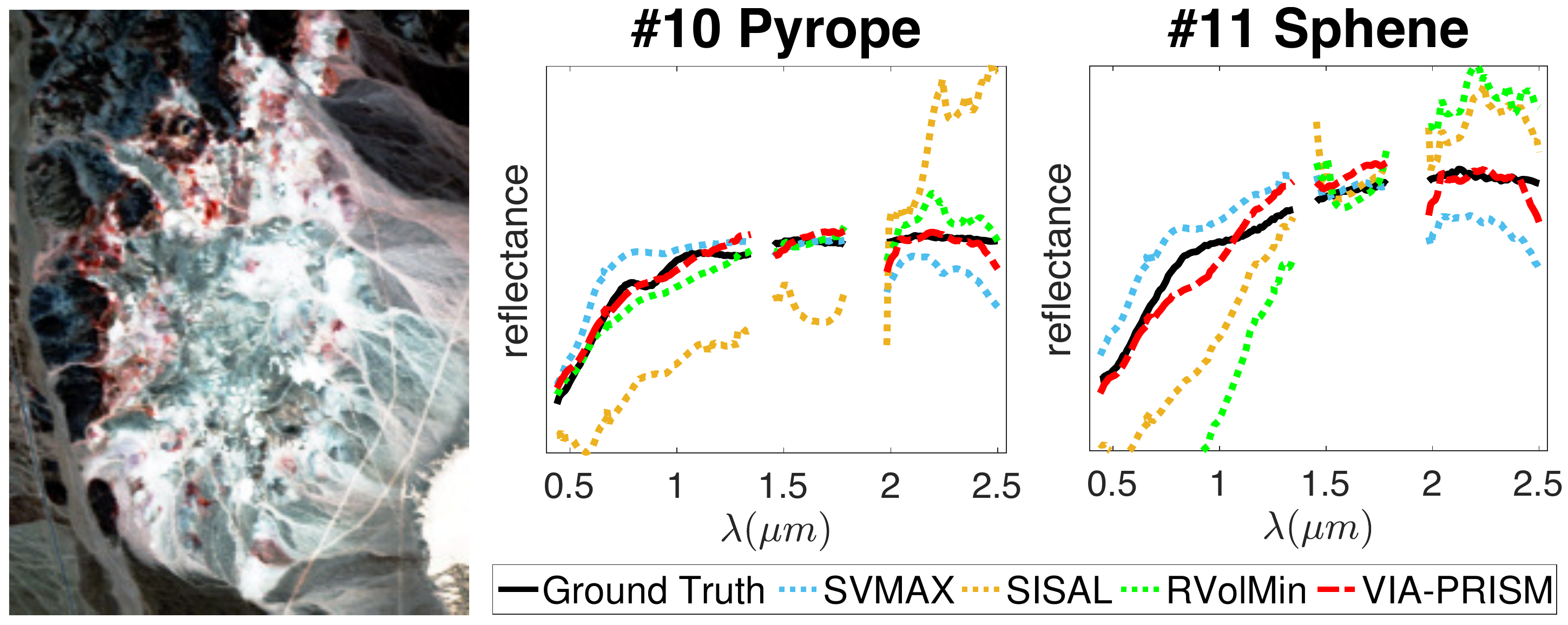}
	\caption{Color composite image from Cuprite Area (left) and the recovered spectra for Pyrope and Sphene (middle and right).}
	\label{fig:cuprite}
\end{figure}

Our experiment setups are as follows.
We take a preprocessed subimage of the Cuprite AVIRIS dataset with $250\times190$ pixels and with $M= 186$ bands; see the left figure in Fig.~\ref{fig:cuprite}.
It is believed that there are $12$ materials (see, e.g., \cite{zhu2017hyperspectral} for a discussion); their names are displayed in Table~\ref{table:performance cuprite}.
As a real-data problem, we do not know the ground truth of the materials' spectral signatures.
As a standard procedure,
the reference spectral signatures from the USGS library~\cite{clark2007usgs} (which records the spectral signatures of numerous materials) corresponding to those 12 materials are used as our believed ground truth.
We measure the performance by the spectral angle distance (SAD) 
$\theta_i(\bA_0, \hat{\bA}) = \cos^{-1} ( \ba_{0,\pi_i}^\top \hat{\ba}_i / (\|{\ba}_{0,\pi_i}\|\| \hat{\ba}_i\|) )$,
where $\bm\pi$ is the solution to $\min_{\bm\pi\in\Pi_N}\frac{1}{N}\sum_{i=1}^{N} \cos^{-1} ( \ba_{0,\pi_i}^\top \hat{\ba}_i / (\|{\ba}_{0,\pi_i}\|\| \hat{\ba}_i\|) )$.

We consider SVMAX, SISAL, RVolMin and VIA-PRISM.
We set $N= 12$.
SISAL has its regularization parameter tuned as $\lambda= 0.001$.
RVolMin has its parameters set as $\lambda = 6$, $p= 1.5$.
We employ an improved version of VIA-PRISM,
wherein we estimate the vertex matrix $\bA$ and the noise variance $\sigma^2$ jointly by incorporating $\sigma^2$ as an extra optimization variable in the ML problem \eqref{eq:ML}.
AM is used to deal with the extended ML problem (see, e.g.,  \cite{wu2019stochastic}, for its ISA counterpart); we shall omit the details here.
Doing so frees us from pre-determining the noise variance $\sigma^2$ when using PRISM.

The SAD performance is listed in Table~\ref{table:performance cuprite}.
We see that all the schemes show good performance in general.
Upon a closer look, SISAL and RVolMin have relatively large estimation errors with ``\#10 Pyrope'' and ``\#11 Sphene''; see Fig.  \ref{fig:cuprite} for the recovered spectral signatures.
We suspect that this may be due to noise sensitivity.
In comparison, VIA-PRISM appears to be less sensitive. 
We should reiterate that there is no parameter tuning with VIA-PRISM.

\begin{table}[h]
	\centering
	\caption{SAD performance on the Cuprite AVIRIS dataset. Blue: the best SAD  among all the tested schemes; red: SADs that are larger than $10$ degrees. }
	\begin{tabular}{l||c|c|c|c}
		\hline\hline
		SAD (degree)        & SVMAX             & SISAL             & RVolMin       & VIA-PRISM
		\\ \hline\hline
		average performance & {\blue \bf 2.89}              & 4.55              & 3.69          & 3.07  \\ \hline\hline
		\#1 Alunite         & {\bf\blue 2.33}   & 2.76              & 3.12          & 3.25       \\
		\#2 Andradite       & 2.97              & 2.61              &{\bf\blue 1.87}& 2.27       \\
		\#3 Buddingtonite   & 3.28              & {\bf\blue 3.21}   & 3.30          & 3.57     \\
		\#4 Dumortierite    & {\bf\blue 2.12}   & 3.11              & 2.32          &4.88       \\
		\#5 Kaolinite$_1 $  & {\bf\blue 2.36}   & 3.38              & 3.51          &4.88              \\
		\#6 Kaolinite$ _2 $ & 2.27              & {\bf\blue 2.24}   & 2.71          &2.79     \\
		\#7 Muscovite       & {\bf\blue 2.49}   & 3.65              & { 3.30}       & 2.50       \\
		\#8 Montmorillonite & 3.39              & 2.39              &{\bf\blue 1.65}&2.09         \\
		\#9 Nontronite      & {\bf\blue 2.07}   & 2.61              & 2.21          & 2.36             \\
		\#10 Pyrope         & 4.07              & {\red 14.78}             & 2.54          & {\bf\blue 1.40}   \\
		\#11 Sphene         & 4.86              & 9.38              & {\red 15.34}         & {\bf\blue 2.28}   \\
		\#12 Chalcedony     & 2.48             & 4.52              & {\blue \bf 2.44}      & 4.53   \\ \hline\hline
	\end{tabular}
	\label{table:performance cuprite}
\end{table}


\section{Conclusion and Discussion}
\label{sect:con}

We studied simplex component analysis under a probabilistic paradigm.
Our study revealed the following results.
\begin{enumerate}[1.]
	\item Our framework, called PRISM, works on an identifiable model, provably.
	Both theory and numerical results suggested that PRISM can leverage large data size to reduce noise.
	
	\item We showed how some powerful state-of-the-art methods, namely, simplex volume minimization (SVMin) methods, are related to PRISM.
	They also appeared to be good approximations of PRISM for the high-SNR regime.
	\item PRISM is a difficult problem; it is an optimization problem that has intractable integrals appearing in the objective function. 
	We studied variational inference approximation (VIA) for PRISM.
	It was shown that VIA-PRISM resembles regularized matrix factorization, and
	there is no parameter to manually tune, unlike some regularized matrix factorization and SVMin methods.
	Our analysis indicated that VIA-PRISM may be a poor approximation under very high SNRs.
	However our numerical results illustrated that VIA-PRISM works well under the medium- or low-SNR regime, and that VIA-PRISM performs better than the state-of-the-art methods for large latent-variable size $N$.
	\item We examined importance sampling approximation (ISA) for PRISM.
	ISA-PRISM is easy to use, and empirically it works very well for small $N$.
	But, computationally, it does not scale well with $N$.
	
\end{enumerate}

As future work we will be interested in efficient realizations of PRISM, e.g., by VIA or by SVMin (improved forms driven by PRISM).
Extension of the current probabilistic model to more complex data models also appears promising.


\appendix

\ifplainver
\section*{Appendix}
\renewcommand{\thesubsection}{\Alph{subsection}}
\else
\section{Appendix}
\fi

\subsection{Proof of Fact~\ref{fact:sim_dist}}
\label{sect:proof fact 2}
By $s_N = 1- \bone^\top \bar{\bs}$, we can write
\[
\bx = \textstyle ( \sum_{i=1}^{N-1} \bb_i s_i ) + \bb_N (1-\sum_{i=1}^{N-1} s_i)  =   \bar{\bB} \bar{\bs} + \bb_N,
\]
where $\bar{\bB}$
is invertible due to the affine independence of $\bB$.
Since the mapping from $\bar{\bs}$ to $\bx$ is bijective,
we can apply transformation of random variables to obtain
\[
p(\bx) = \frac{1}{|\det(\bar{\bB})|} D( \bar{\bB}^{-1}(\bx - \bb_N); \bone),
\]
where
$D$ is defined in \eqref{eq:Dir_rig}.
It can be verified that
\[
\bar{\bB}^{-1}(\bx - \bb_N) \in \tilde{\Delta} ~ \Longleftrightarrow \bx \in \bconv(\bB),
\]
and hence $D( \bar{\bB}^{-1}(\bx - \bb_N); \bone) = (N-1)! \cdot \indfn{\bconv(\bB)}(\bx)$.
Also, it can seen from \eqref{eq:svol} that ${\rm vol}(\bB) = |\det(\bar{\bB})|/(N-1)!$ when $\bar{\bB}$ is square. The proof is complete.

\subsection{The Formal Proof of ML Identifiability}
\label{sect:formal identifiability}	
The formal proof of Theorem~\ref{thm:id} is as follows.
Consider the following propositions, whose proofs are shown in Appendix \ref{sect:app_prop:1} and \ref{sect:app_prop:2}.
\begin{Prop} \label{prop:1}
	If $\bA$ is such that $p(\by; \bA) = p(\by; \bA_0)$ for all $\by$, then
	(a) $\aff(\bA) = \aff(\bA_0)$; (b) $\svol(\bA) = \svol(\bA_0)$.
\end{Prop}
\begin{Prop} \label{prop:2}
	Let $(\bQ,\bd) \in \Rbb^{M \times (N-1)} \times \Rbb^M$, with $\bQ$ being semi-orthogonal (i.e., $\bQ^\top \bQ = \bI$), such that $\aff(\bA_0) = \sspan(\bQ) + \bd$;
	such $(\bQ,\bd)$ exists.
	Let
	\[
	q(\bz;\bA) = (\sqrt{2\pi} \sigma)^{M-N+1} p(\by= \bQ\bz + \bd; \bA), ~ \bz \in \Rbb^{N-1},
	\]
	where, on the right-hand side of the above equation, $p(\by;\bA)$  is given in \eqref{eq:pyA}.
	If $\bA$ is such that $\aff(\bA) = \aff(\bA_0)$, then $q(\bz;\bA)$ takes the same form as \eqref{eq:pyA_sim}, specifically,
	\beq \label{eq:qzB}
	q(\bz; \bA) = \int_{\Rbb^{N-1}} \varphi_\sigma(\bz - \bx) p(\bx; \bB) {\rm d}\bx := q(\bz;\bB),
	\eeq
	where
	$\bB \in \Rbb^{(N-1) \times N}$ is affinely independent and satisfies $\bA = \bQ\bB + \bd \bone^\top$;
	$p(\bx;\bB)$ is
	defined in \eqref{eq:pxA}.
\end{Prop}

Proposition~\ref{prop:1} is obtained by observing the first- and second-order moments of $\by \sim p( \cdot ;\bA)$.
Proposition~\ref{prop:2} is a dimensionality reduction result which is reminiscent of \cite[Lemma~2]{Chan2009} in spirit.
Invoking Propositions~\ref{prop:1}.(a) and \ref{prop:2}, we have
\begin{align}
	& p(\by; \bA) = p(\by; \bA_0) \text{~ for all $\by$} 
	\quad  \Longrightarrow \quad 
	q(\bz; \bB) = q(\bz; \bB_0) \text{~ for all $\bz$}, \label{eq:form_proof0}
\end{align}
where $\bB$ and $\bB_0$, defined by the way as in Proposition~\ref{prop:2}, are affinely independent.
As $q(\bz; \bB)$ (cf. \eqref{eq:qzB}) takes the same form as $p(\by;\bA)$ in \eqref{eq:pyA_sim} in the intuitive proof in Section \ref{sect:proof_intuit}, most of the steps in the intuitive proof apply.
The exception is the implication
\begin{align}
	& \check{p}(\bxi;\bB) = \check{p}(\bxi;\bB_0) \text{~ for all $\bxi$} 
	\quad 
	\Longrightarrow 
	\quad 
	p(\bx; \bB) =  p(\bx; \bB_0) \text{~ for all $\bx$}. \label{eq:form_proof1}
\end{align}

We complete the proof by confirming the implication \eqref{eq:form_proof1}.
Consider the FT result below.
\begin{Fact}{\cite[Theorem~1.16]{stein2016introduction}} \label{fact:elias}
	If $f: \Rbb^n \rightarrow \Rbb$ belongs to $L^1(\Rbb^n)$, the space of all measurable functions defined on $\Rbb^n$ and with $\int_{\Rbb^n} \| f(\bx) \|_1 {\rm d}\bx < \infty$ ($\| \cdot \|_1$ denotes the $1$-norm), then
	\[
	\int_{\Rbb^n} \check{\varphi}_{\varepsilon}(\bxi) \check{f}(\bxi) e^{j 2 \pi \bxi^\top \bz} {\rm d}\bxi =
	\int_{\Rbb^n} {\varphi}_{\varepsilon}(\bz - \bx) f(\bx)  {\rm d}\bx
	\]
	for all $\varepsilon > 0$.
\end{Fact}
Applying Fact~\ref{fact:elias} to both sides of \eqref{eq:form_proof1} gives
\beq \label{eq:form_proof2}
\int_{\Rbb^{N-1}} {\varphi}_{\varepsilon}(\bz - \bx) p(\bx;\bB)  {\rm d}\bx
= \int_{\Rbb^{N-1}} {\varphi}_{\varepsilon}(\bz - \bx) p(\bx;\bB_0)  {\rm d}\bx.
\eeq
Also, we apply  Proposition~\ref{prop:1}.(b) to \eqref{eq:form_proof0} (with $\bA=\bB$) to get $\svol(\bB) = \svol(\bB_0)$.
Eq.~\eqref{eq:form_proof2} can thus be reduced to
\[
\int_{\bconv(\bB)} {\varphi}_{\varepsilon}(\bz - \bx)  {\rm d}\bx
= \int_{\bconv(\bB_0)} {\varphi}_{\varepsilon}(\bz - \bx)  {\rm d}\bx.
\]
It can be shown that, for an open convex set $\setC$, we have\footnote{Hint: for the uppercase of \eqref{eq:form_proof3}, note that $\bz \in \setC$ is an interior point of $\setC$.
	For the lowercase, consider $\int_{\mathcal{H}} {\varphi}_{\varepsilon}(\bz - \bx)  {\rm d}\bx$ where $\mathcal{H}$, $\setC \subset \mathcal{H}$, is a halfspace associated with the separating hyperplane of $\bz$ and $\setC$.}
\beq \label{eq:form_proof3}
\lim_{\varepsilon \rightarrow 0} \int_{\setC} {\varphi}_{\varepsilon}(\bz - \bx)  {\rm d}\bx
\left\{ \begin{array}{ll} = 1, & \bz \in \setC,  \\
	\leq 1/2, & \bz \notin \setC.
\end{array} \right.
\eeq 
The above two equations lead to
\[
\bz \in \bconv(\bB) \quad  \Longleftrightarrow \quad  \bz  \in \bconv(\bB_0),
\]
and subsequently, $p(\bx; \bB) =  p(\bx; \bB_0)$. The proof is complete.

\subsection{Proof of Proposition~\ref{prop:1}}
\label{sect:app_prop:1}

First, consider the mean and covariance of $\by$ for a given $\bA$.
Recall from the model in Section~\ref{sect:formulation} of the main manuscript that $\by = \bA \bs+ \bv$, $\bs \sim D(\cdot;\bone)$, $\bv \sim \varphi_\sigma$, with $\bv$ independent of $\bs$.
Using Fact~\ref{fac:Dir_moments}, we get
\begin{align}
	\Exp[ \by ] & = \tfrac{1}{N} \bA \bone,  \label{eq:proofprop1_mean} \\
	\bC_\by & := \cov(\by) = \bA \bC_{\bs} \bA^\top + \sigma^2 \bI,  \nonumber  \\
	\bC_\bs & := \cov(\bs) = \tfrac{1}{(N+1)N} \left( \bI - \tfrac{1}{N} \bone \bone^\top \right). \nonumber
\end{align}
Let $\bU \in \Rbb^{N \times (N-1)}$ be any semi-orthogonal matrix such that $\bU^\top \bone = \bzero$.
It can be verified that $\bU \bU^\top = \bI - \bone \bone^\top / N$.
The covariance $\bC_\by$ can thus be simplified to
\beq \label{eq:proofprop1_cov}
\bC_\by = \tfrac{1}{(N+1)N} (\bA\bU)(\bA\bU)^\top + \sigma^2 \bI.
\eeq

Second, suppose $p(\by;\bA) = p(\by;\bA_0)$ for all $\by$.
We see from \eqref{eq:proofprop1_mean} and \eqref{eq:proofprop1_cov} that
\begin{align}
	\tfrac{1}{N} \bA \bone & = \tfrac{1}{N} \bA_0 \bone, \label{eq:proofprop1_mean_eq} \\
	(\bA\bU)(\bA\bU)^\top & = (\bA_0 \bU)(\bA_0\bU)^\top. \label{eq:proofprop1_cov_eq}
\end{align}
Consider the following results.
\begin{Lemma} \label{lem:aff}
	Let $\bA \in \Rbb^{m \times n}$ be any matrix.
	Let $\bU \in \Rbb^{n \times (n-1)}$ be a semi-orthogonal matrix such that $\bU^\top \bone = \bzero$. Then
	\begin{enumerate}[(a)]
		\item $\aff(\bA) = \sspan(\bar{\bA}) + \bd$ for any $\bd \in \aff(\bA)$;
		\item $\sspan(\bar{\bA}) = \sspan(\bA\bU)$;
		\item $\det(\bar{\bA}^\top \bar{\bA} ) = C \det((\bA\bU)^\top (\bA\bU) )$ for some constant $C$.
	\end{enumerate}
\end{Lemma}
The proof of Lemma~\ref{lem:aff} will be shown in Appendix \ref{sect:lem:aff}.
We will also need the following results.
\begin{Lemma} \label{lem:mat}
	Let $\bF, \bG \in \Rbb^{m \times n}$, $m \geq n$.
	If $\bF \bF^\top = \bG \bG^\top$,
	\begin{enumerate}[(a)]
		\item $\sspan(\bF) = \sspan(\bG)$;
		\item $\det(\bF^\top \bF) = \det(\bG^\top \bG)$.
	\end{enumerate}
\end{Lemma}
We omit the proof of Lemma~\ref{lem:mat} as it is basic in matrix analysis (hint: use singular value decomposition).
By Lemma~\ref{lem:aff}.(a)--(b) we have
\beq
\begin{aligned}
	\aff(\bA) & = \sspan(\bA\bU) + \tfrac{1}{N} \bA \bone, \\
	\aff(\bA_0) & = \sspan(\bA_0 \bU) + \tfrac{1}{N} \bA_0 \bone.
\end{aligned}
\eeq
By \eqref{eq:proofprop1_cov_eq}
and Lemma~\ref{lem:mat}.(a), we have $\sspan(\bA\bU) = \sspan(\bA_0 \bU)$.
In addition, since $\tfrac{1}{N} \bA \bone  = \tfrac{1}{N} \bA_0 \bone$ (cf. \eqref{eq:proofprop1_mean_eq}), we have $\aff(\bA) = \aff(\bA_0)$.
Moreover, from \eqref{eq:proofprop1_cov_eq}, Lemma~\ref{lem:aff}.(c) and Lemma~\ref{lem:mat}.(b), we obtain $\det(\bar{\bA}^\top \bar{\bA}) = \det(\bar{\bA}^\top_0 \bar{\bA}_0)$.
This leads to $\svol(\bA) = \svol(\bA_0)$.
The proof is complete.

\subsection{Proof of Lemma~\ref{lem:aff}}
\label{sect:lem:aff}

Lemma~\ref{lem:aff}.(a) is basic.
Concisely, from the definition of $\aff$, it is immediate that $\aff(\bA) = \sspan(\bar{\bA}) + \ba_n$.
As $\bd \in \aff(\bA)$, we can write $\bd = \bb + \ba_n$ for some $\bb \in \sspan(\bar{\bA})$.
As $\sspan(\bar{\bA}) + \bb = \sspan(\bar{\bA})$, we have Lemma~\ref{lem:aff}.(a).

For Lemma~\ref{lem:aff}.(b) it suffices to show $\sspan(\bU) = \sspan(\bar{\bI})$.
By noting $\bar{\bI} = [~ \be_1 - \be_n, \ldots, \be_{n-1} - \be_n ~]$,
we see that
\begin{align*}
	\bx \in \sspan(\bar{\bI}) ~ \Longleftrightarrow ~ & \bx = ( x_1,\ldots,x_{n-1}, \textstyle{- \sum_{i=1}^{n-1} x_i} ) \\
	\Longleftrightarrow ~   & \bx \in \setN(\bone^\top) := \{ \bx \in  \Rbb^n \mid \bone^\top \bx = 0 \}.
\end{align*}
The set $\setN(\bone^\top)$ equals the orthogonal complement of $\sspan(\bone)$, which is $\sspan(\bU)$. The proof is done.

For Lemma~\ref{lem:aff}.(c), we have the following.
Since $\sspan(\bU) = \sspan(\bar{\bI})$ in the preceding proof, we can write $\bar{\bI} = \bU \bm \Theta$ for some $\bm \Theta \in \Rbb^{(n-1) \times (n-1)}$.
By equaling $\bar{\bA} = \bA \bar{\bI} = \bA \bU \bm \Theta$,
\begin{align*}
	\det(\bar{\bA}^\top \bar{\bA}) & = \det(  \bm \Theta^\top ( (\bA \bU)^\top (\bA\bU) )  \bm \Theta ) \\
	& = (\det(\bm \Theta))^2 \det((\bA \bU)^\top (\bA\bU)),
\end{align*}
where we exploit the fact that $\bm \Theta$ and $(\bA \bU)^\top (\bA\bU)$ are square.
Letting $C = (\det(\bm \Theta))^2$ completes the proof.

\subsection{Proof of Proposition~\ref{prop:2}}
\label{sect:app_prop:2}

First it is clear from Lemma~\ref{lem:aff}.(a) that we can write $\aff(\bA_0) = \sspan(\bQ) + \bd;$
here $\bQ$ is an orthogonal basis for $\sspan(\bar{\bA}_0)$, and $\bQ$ has $N-1$ columns because $\bar{\bA}_0$ has full-column rank.
The identity $\aff(\bA) = \aff(\bA_0)$ implies that i) $\bA$ is affinely independent, and ii)
we can write $\ba_i = \bQ \bb_i + \bd$ for some coefficient $\bb_i \in \Rbb^{N-1}$. This leads to
\[
\bA = \bQ \bB + \bd \bone^\top  ~ \Longleftrightarrow ~ \bB = \bQ^\top (\bA - \bd \bone^\top).
\]
Also, $\bB$ must be affinely independent, for otherwise $\bA$ will be affinely dependent as one may verify.

Second, consider $\| \by - \bA\bs \|^2$ for $\by = \bQ \bz + \bd$, $\bs \in \Delta$:
\begin{align*}
	\| \by - \bA \bs \|^2 & = \| \bQ \bz + \bd -  ( \bQ \bB + \bd \bone^\top ) \bs \|^2 \\
	& = \| \bQ(\bz - \bB \bs) \|^2 = \| \bz - \bB \bs \|^2,
\end{align*}
where we have used $\bone^\top \bs = 1$.
Putting the above equality to the PDF $p(\by;\bA)$ in \eqref{eq:pyA}, we obtain
\begin{align*}
	q(\bz;\bA) & = (\sqrt{2\pi} \sigma)^{M-N+1} p(\by= \bQ\bz+ \bd;\bA) \\
	& =  (N-1)! \int \varphi_\sigma(\bz - \bB\bs) \indfn{\bar{\Delta}}(\bs) {\rm d}\mu(\bs).
\end{align*}
By the change of variables $\bx = \bB\bs = \bar{\bB} \bar{\bs} + \bb_N$, one will find that
\[
q(\bz;\bA) = \frac{(N-1)!}{\det(\bar{\bB)}} \int_{\Rbb^{N-1}} \varphi_\sigma(\bz - \bx) \indfn{\bconv(\bB)}(\bx) {\rm d}\bx;
\]
the proof is identical to that in Fact~\ref{fact:sim_dist}.
The proof is complete.

\subsection{Dimensionality Reduction}
\label{sect:app_DR}

Here we describe the dimensionality reduction (DR) procedure for \SCA\ and explain why it works.

\begin{algorithm}[htb!]
	\caption{DR for \SCA} \label{alg:dr}
	\begin{algorithmic}[1]
		\State {\bf given} a collection of data points $\by_1,\ldots,\by_T \in \Rbb^M$ and the model order $N$
		\State compute the sample mean $\hat{\bmu}_y = ( \sum_{t=1}^T \by_t )/T$
		\State compute the sample covariance $\hat{\bC}_ y = ( \sum_{t=1}^T (\by_t  - \hat{\bmu}_y )( \by_t  - \hat{\bmu}_y )^\top )/T$
		\State compute the eigenvectors of $\hat{\bC}_\by$ associated with the first $N-1$ principal eigenvalues, store them in  $\hat{\bQ} \in \Rbb^{M \times (N-1)}$
		\State apply the DR
		\beq \label{eq:dr_routine}
		\bz_t  = \hat{\bQ}^\top (\by_t  - \hat{\bmu}_y ), \quad t=1,\ldots,T,
		\eeq
		where $\bz_t \in \Rbb^{N-1}$ is a dimension-reduced point of $\by_t$
		\State {\bf output} $\bz_1,\ldots\bz_T$
	\end{algorithmic}
\end{algorithm}

We begin by posting the DR procedure in Algorithm \ref{alg:dr}.
It is the same as the standard PCA, 
but there is a difference with its result and explanation in the context of \SCA.
First, we describe the result.
Recall the \SCA\ model 
\beq \label{eq:model_dr}
\by_t  = \bA_0 \bs_t + \bv_t.
\eeq 
We argue that the dimension-reduced points $\bz_t$'s outputted by Algorithm \ref{alg:dr} can be modeled as
\beq \label{eq:dr_model}
\bz_t = \bB_0 \bs_t + \bw_t,
\eeq
where $\bB_0 \in \Rbb^{(N-1) \times N}$ is affinely independent and should satisfy the relation
\beq \label{eq:dr_relate}
\bA_0 = \hat{\bQ} \bB_0 + \hat{\bmu}_y \bone^\top;
\eeq 
$\bw_ t$ is Gaussian noise with mean $\bzero$ and covariance $\sigma^2 \bI$.
Since \eqref{eq:dr_model} takes the same form as \eqref{eq:model_dr}, we can perform \SCA\ by
i) estimating $\bB_0$ from the dimension-reduced points $\bz_1,\ldots,\bz_T$ via a \SCA\  algorithm, and then
ii) forming an estimated $\bA_0$ from the estimated $\bB_0$ via  \eqref{eq:dr_relate}.

Second, we explain why the above result holds.
According to the derivations in Appendix~\ref{sect:app_prop:1},
the mean and covariance of $\by_t$ under  \eqref{eq:model_dr} are
\begin{align}
	\bmu_y & := \Exp[ \by_t ] = \frac{1}{N} \bA_0 \bone, \nonumber \\
	\bC_y & := \cov(\by_t)  = \tfrac{1}{(N+1)N} (\bA_0 \bU) (\bA_0 \bU)^\top  + \sigma^2 \bI, \label{eq:Cy}
\end{align}
for a semi-orthogonal $\bU \in \Rbb^{N \times (N-1)}$ such that $\bU^\top \bone = \bzero$.
Let $\bQ \in \Rbb^{M \times (N-1)}$
be a matrix whose columns are the eigenvectors of $\bC_y$ associated with the first $N-1$ principal eigenvalues.
It can be shown from \eqref{eq:Cy} that $\sspan(\bQ) = \sspan(\bA_0 \bU) = \sspan(\bar{\bA}_0)$;
note that we use the assumption that $\bA_0$ is affinely independent.
Moreover, as a corollary of the derivations in Appendix \ref{sect:app_prop:2},
$\bA_0$ can be expressed as
\beq \label{eq:AB_equiv}
\bA_0 = \bQ \bB_0 + \bmu_y \bone^\top  ~ \Longleftrightarrow ~ \bB_0 = \bQ^\top (\bA - \bmu_y \bone^\top),
\eeq
with $\bB_0$ being affinely independent.
Applying the left-hand side of \eqref{eq:AB_equiv} to \eqref{eq:dr_routine} yields the dimension-reduced data model \eqref{eq:dr_model}.
Note that Algorithm \ref{alg:dr} uses sample mean and sample covariance to do the above DR task.
Our explanation is done.

As a minor remark, the above explanation is slightly new.
The prior study \cite{Chan2009,Ma2014HU} derived Algorithm \ref{alg:dr}  by using the deterministic CG notion, which assumes no noise.
The above explanation is statistical and assumes the presence of noise.

\subsection{Further Discussion with the SVMin-PRISM Relationships}
\label{sect:discuss_svmin}

We discuss two issues arising from the SVMin-PRISM relationships shown in Section~\ref{sect:relate} of the main manuscript.
The first issue is that we focused on the case of $M=N-1$,
and one may ask whether this can be lifted to $M \geq N-1$.
We give two answers, one simple and one complicated.
The simple one is no, but the issue can be easily sidestepped.
Our derivations exploit friendly properties of full-dimensional simplices,
and that restricts us to $M= N-1$.
But it is well-known in CG that we can handle the issue by dimensionality reduction  \cite{Chan2009,Ma2014HU}; e.g., the one in Appendix \ref{sect:app_DR}.
The complicated answer is that it is possible, but the result will not be simple.
In Appendix~\ref{sect:app_ESA}, we show that the extension of the edge-smooth approximation in Section~\ref{sect:connect_SVMin_SSMF} to $M \geq N-1$ gives rise to an approximate ML
\ifconfver
\begin{align*}
	\min_{\bA \in \setA, \bS \in \Delta^T} &
	\log \svol(\bA) +  \tfrac{1}{\lambda T}  \| \bY - \bA \bS \|^2   \\
	& + \tfrac{1}{T}
	\left( \tfrac{1}{2\sigma^2 } -  \tfrac{1}{\lambda}  \right) \min_{ \bXi^\top \bone = \bone } \| \bY - \bA \bXi \|^2.
\end{align*}
\else
\beq
\min_{\bA \in \setA, \bS \in \Delta^T}
\log \svol(\bA) +  \tfrac{1}{\lambda T}  \| \bY - \bA \bS \|^2 + \tfrac{1}{T}
\left( \tfrac{1}{2\sigma^2 } -  \tfrac{1}{\lambda}  \right) \min_{ \bXi^\top \bone = \bone } \| \bY - \bA \bXi \|^2.
\nonumber
\eeq
\fi
We see that the above problem resembles the SVMin-SSMF \eqref{eq:VRMF}, but it has an additional term (the third term).

The second issue we want to discuss is with the affine independence constraint; specifically, $\bA \in \setA$ or $\bB \in \setB$ in \eqref{eq:MVES_from_ML}, \eqref{eq:VRMF}, \eqref{eq:SISAL}, \eqref{eq:aml_prob}, etc.
This constraint is necessary in order to be mathematically correct in our development,
but it is often discarded for ease of realization in practice.
It is generally safe to remove the affine independence constraint from the original noiseless SVMin \eqref{eq:MVES_from_ML}, SISAL \eqref{eq:SISAL} and the chance-promoting SVMin \eqref{eq:aml_prob};
e.g., if the $\by_t$'s are well spread such that any data enclosing simplex $\conv(\bA)$ has to be full-dimensional, then the affine independence of $\bA$ will be automatically  satisfied for \eqref{eq:MVES_from_ML}.
For SVMin-SSMF \eqref{eq:VRMF}, its objective value approaches $-\infty$ as $\bA$ approaches affine dependence.
The trick to get this around is to replace  the simplex volume \eqref{eq:svol} by $(\det( \bar{\bA}^\top \bar{\bA} + \varepsilon \bI)))^{1/2} /((N-1)!)$ for some small $\varepsilon > 0$ \cite{fu2016robust}.

\subsection{Edge-Smooth Approximation for $M \geq N-1$}
\label{sect:app_ESA}

Here we extend the edge-smooth approximation in Section~\ref{sect:connect_SVMin_SSMF} of the main manuscript to the more general case of $M \geq N-1$.
First we reformulate $p(\by;\bA)$ in \eqref{eq:pyA} such that we can apply edge-smooth approximation.
Let $\bA$ be affinely independent.
According to the proof of Proposition~\ref{prop:2}, we know that i) the affine hull of $\bA$ can be characterized as $\aff(\bA) = \sspan(\bQ) + \bd$ for some semi-orthogonal $\bQ \in \Rbb^{M \times (N-1)}$ and for some $\bd$; ii) $\bA$ can be characterized as $\bA = \bQ \bB + \bd \bone^\top$ for some affinely independent $\bB \in \Rbb^{(N-1) \times N}$.
Let $\tilde{\bQ} \in \Rbb^{M \times (M-N+1)}$ be such that $\bU = [~ \bQ ~ \tilde{\bQ} ~]$ is orthogonal.
For any $\bs \in \Delta$ we have
\begin{align}
	\| \by - \bA \bs \|^2 & = \| \bU^\top [ \by - (\bQ \bB + \bd \bone^\top) \bs ] \|^2
	\nonumber \\
	& = \| \bz_1 - \bB \bs \|^2 + \| \bz_2 \|^2, \label{eq:ESA_t1}
\end{align}
where $\bz_1$ and $\bz_2$ are defined as
\[
\begin{bmatrix}
	\bz_1 \\ \bz_2
\end{bmatrix}
=
\begin{bmatrix}
	\bQ^\top (\by - \bd)  \\ \tilde{\bQ}^\top (\by - \bd)
\end{bmatrix}
= \bU^\top (\by - \bd).
\]
With \eqref{eq:ESA_t1}, we can write $\varphi_\sigma(\by- \bA \bs) = \varphi_\sigma(\bz_2) \cdot \varphi_\sigma(\bz_1 - \bB \bs )$.
The PDF $p(\by;\bA)$ in \eqref{eq:pyA} can be rewritten as
\begin{align}
	p(\by;\bA) & = \varphi_\sigma(\bz_2) \cdot \int \varphi_\sigma(\bz_1 - \bB \bs ) p(\bs) {\rm d} \mu(\bs) \label{eq:ESA_t2} \\
	& = \varphi_\sigma(\bz_2) \cdot \int_{\Rbb^{N-1}} \varphi_\sigma(\bz_1 - \bx ) p(\bx;\bB) {\rm d}\bx, \label{eq:ESA_t3}
\end{align}
where $p(\bx;\bB)$ is given in \eqref{eq:pxA}; here we transform \eqref{eq:ESA_t2} to \eqref{eq:ESA_t3} by the same formulation as in \eqref{eq:pyA_sim}.

Second we apply the same edge-smooth approximation in \eqref{eq:e_smooth} to the integral \eqref{eq:ESA_t3}; i.e.,
\[
\int_{\Rbb^{N-1}} \varphi_\sigma(\bz_1 - \bx ) p(\bx;\bB) {\rm d}\bx
\approx \frac{1}{\svol(\bB)} e^{-  \frac{1}{\lambda} \cdot {\rm dist}(\bz_1, \conv(\bB) )^2}.
\]
This results in
\beq  \label{eq:ESA_t3_1}
- \log p(\by;\bA) \approx \frac{1}{2\sigma^2} \| \bz_2 \|^2 + \log \svol(\bB) +  \frac{1}{\lambda} \cdot  {\rm dist}(\bz_1, \conv(\bB) )^2.
\eeq
Now, observe that
\begin{align}
	{\rm dist}(\by, \conv(\bA) )^2  & = \min_{\bs \in \Delta} \| \by - \bA \bs \|^2  \nonumber \\
	& = \min_{\bs \in \Delta} \| \bz_1 - \bB \bs \|^2 + \| \bz_2 \|^2  \nonumber \\
	& = {\rm dist}(\bz_1,\conv(\bB))^2 + \| \bz_2 \|^2
	\label{eq:ESA_t4}  \\
	{\rm dist}(\by, \aff(\bA) )^2 & = \min_{\bone^\top \bs = 1} \| \by - \bA \bs \|^2  \nonumber \\
	& = \min_{\bone^\top \bs = 1} \| \bz_1 - \bB \bs \|^2 + \| \bz_2 \|^2 \nonumber \\
	& = \min_{\bar{\bs} \in \Rbb^{N-1}} \| \bz_1 - \bb_N - \bar{\bB} \bar{\bs} \|^2 + \| \bz_2 \|^2 \nonumber \\
	& = \| \bz_2 \|^2, \label{eq:ESA_t5}  \\
	\det(\bar{\bA}^\top \bar{\bA} ) &
	= \det( (\bQ \bar{\bB})^\top (\bQ \bar{\bB}) )
	= \det( \bar{\bB}^\top \bar{\bB} ),   \label{eq:ESA_t6}
\end{align}
where
\eqref{eq:ESA_t4} is due to \eqref{eq:ESA_t1};
\eqref{eq:ESA_t5} is due to \eqref{eq:ESA_t1} and the fact that  $\bar{\bB}$ is invertible;
\eqref{eq:ESA_t6} is due to $\bar{\ba}_i = \ba_i  - \ba_N= \bQ (\bb_i - \bb_N) = \bQ \bar{\bb}_i$.
Substituting \eqref{eq:ESA_t4}--\eqref{eq:ESA_t6} into \eqref{eq:ESA_t3_1},
and then applying it to the ML problem \eqref{eq:ML}, we obtain
the edge-smooth approximation of the ML problem as
\ifconfver
\begin{align*}
	\min_{\bA \in \setA, \bS \in \Delta^T} &
	\log \svol(\bA) +  \tfrac{1}{\lambda T}   \| \bY - \bA \bS \|^2   \\
	& + \tfrac{1}{T}
	\left( \tfrac{1}{2\sigma^2 } -  \tfrac{1}{\lambda} \right) \min_{ \bXi^\top \bone = \bone } \| \bY - \bA \bXi \|^2
\end{align*}
\else
\beq
\min_{\bA \in \setA, \bS \in \Delta^T}
\log \svol(\bA) +   \tfrac{1}{\lambda T}  \| \bY - \bA \bS \|^2 + \tfrac{1}{T}
\left( \tfrac{1}{2\sigma^2 } -  \frac{1}{\lambda} \right) \min_{ \bXi^\top \bone = \bone } \| \bY - \bA \bXi \|^2.
\nonumber
\eeq
\fi

\subsection{Proof of Proposition~\ref{prop:cvx_VIA_sub}}
\label{sect:app_cvx_VIA}

We will assume $\alpha > 0$ without mentioning.
By the expansion
\[
\log \Gamma(\alpha) = -\gamma \alpha - \log \alpha + \sum_{k=1}^\infty \left[ \frac{\alpha}{k} - \log\left( 1 + \frac{\alpha}{k} \right) \right],
\]
where $\gamma$ is the Euler–Mascheroni constant
(see, e.g., \cite[p.~204]{boros2004irresistible}), one can show that the function $h$ in \eqref{eq:fn_h} has its double differentiation given by
\[
h''(\alpha)  = \sum_{k=0}^\infty \left[ \frac{1}{(\alpha+k)^2} - \frac{2(\alpha-1)}{(\alpha+k)^3} \right].
\]
We will show that
\[
h''(\alpha) \geq \frac{1}{(n+ \alpha)^2} > 0, \quad n =  \max\{ 0, \lceil 2\alpha - 3 \rceil \},
\]
and thereby confirms the strict convexity of $h$
(here $\lceil x \rceil$ denotes the ceiling of $x$).
To facilitate, let
\[
f_\alpha(y) = \frac{1}{(\alpha+y)^2} - \frac{2(\alpha-1)}{(\alpha+y)^3}
\]
and write
\[
h''(\alpha) = \underbrace{\sum_{k=0}^{n-1} f_\alpha(k)}_{:= p(\alpha)} + \underbrace{\sum_{k=n}^{\infty} f_\alpha(k)}_{:= q(\alpha)}.
\]
Our task is to derive lower bounds of $p$ and $q$.

First we deal with $q$.
From
\begin{align*}
	f'_\alpha(y) & = - \frac{2}{(\alpha + y)^3} + \frac{6(\alpha-1)}{(\alpha + y)^4}
	=
	-\frac{2}{(\alpha + y)^4} ( y - 2\alpha + 3),
\end{align*}
we see that $f_\alpha(y)$ is  nonincreasing for $y \geq n$ (note $n \geq 2\alpha - 3$).
This implies that, for $k \geq n$, we have $f_\alpha(k) \geq f_\alpha(y)$ for any $y \in [k,k+1]$.
We can therefore write
\begin{align}
	q(\alpha) & \geq \sum_{k=n}^\infty \int_{k}^{k+1} f_\alpha(y) {\rm d}y = \int_n^\infty f_\alpha(y) {\rm d}y \nonumber \\
	& = \frac{1}{n+ \alpha} - \frac{\alpha-1}{(n+\alpha)^2} = \frac{n+1}{(n+\alpha)^2}.
	\label{eq:q_bound}
\end{align}

Second we handle $p$.
Let $p_m(\alpha) = \sum_{k=0}^{m-1} f_\alpha(k)$, $m \geq 1$.
We claim that
\beq \label{eq:pm}
p_m(\alpha) \geq - \frac{m}{(m+\alpha)^2},
\eeq
and consequently, $p(\alpha) = p_n(\alpha) = - n/(n+\alpha)^2$.
To show it we use the following inequality
\beq \label{eq:a_lemma}
- \frac{1}{(m+\alpha)^2} + \frac{2}{(m+\alpha)^3} \geq  - \frac{1}{(m+1+\alpha)^2}
\eeq
for any integer $m \geq 1$. The proof of \eqref{eq:a_lemma} will be provided later.
Using \eqref{eq:a_lemma} with $m=0$, we observe that
\[
p_1(\alpha) = f_\alpha(0) = -\frac{1}{\alpha^2} + \frac{2}{\alpha^3} \geq - \frac{1}{(1+\alpha)^2},
\]
which is \eqref{eq:pm} for the case of $m=1$.
To show the other cases, suppose  $p_m(\alpha) \geq - m/(m+\alpha)^2$ is true. Then
\begin{align*}
	p_{m+1}(\alpha) & = p_m(\alpha) + f_\alpha(m) \\
	& \geq -\frac{m}{(m+\alpha)^2} + \frac{1}{(m+\alpha)^2} - \frac{2(\alpha-1)}{(m+\alpha)^3} \\
	& = (m+1) \left[ -\frac{1}{(m+\alpha)^2} + \frac{2}{(m+\alpha)^3} \right] \\
	& \geq - (m+1) \frac{1}{(m+1+\alpha)^2},
\end{align*}
where the last inequality is due to \eqref{eq:a_lemma}.
Hence, by induction, we confirm that \eqref{eq:pm} is true for all $m \geq 1$.

Third, by \eqref{eq:q_bound} and \eqref{eq:pm}, we obtain the desired result $h''(\alpha) \geq 1/(n+\alpha)^2$.
Before we finish, we should provide the proof of \eqref{eq:a_lemma}.
Let $g(y) = 1/(1+y)^2$. Since $g$ is convex on $\Rbb_+$, by the first-order condition of convex functions, i.e., $g(y) \geq g(x) + g'(x)(y-x)$ for $x,y \in \Rbb_+$, we have
\begin{align*}
	& g(\alpha+ m) \geq g(\alpha+m-1) + g'(\alpha+m-1)  \\
	\Longleftrightarrow \quad & \frac{1}{(\alpha+m+1)^2} \geq \frac{1}{(\alpha+m)^2} - \frac{2}{(\alpha+m)^3}.
\end{align*}
The proof is complete.

\subsection{An ADMM Algorithm for Problem \eqref{eq:cvx_VIA_sub}}
\label{sect:app_cvx_VIA_admm}

Here we design an ADMM algorithm for solving the convex problem  \eqref{eq:cvx_VIA_sub} efficiently.
To reduce notational overheads, let $g(\balp)= g(\bA,\balp,\eta;\by)$.
We reformulate problem \eqref{eq:cvx_VIA_sub} as
\beq \label{eq:prob_admm}
\begin{aligned}
	\min_{\balp, \bbeta \in \Rbb^N} & \, g(\balp) + \textstyle \sum_{i=1}^N h(\beta_i) \\
	{\rm s.t.} & \, \bone^\top \balp = \eta, ~ \bbeta > \bzero, ~ \balp= \bbeta,
\end{aligned}
\eeq
where we split the variable into two, one for $g$ and another for $h$.
Denote the augmented Lagrangian of
the above problem
by
\beq
L_\rho(\balp,\bbeta,\blam) = g(\balp) + \textstyle \sum_{i=1}^N h(\beta_i) + \blam^\top( \balp - \bbeta) + \tfrac{\rho}{2} \| \balp - \bbeta \|^2,
\nonumber
\eeq
where $\rho > 0$ is given.
Following the ADMM literature \cite{boyd2011distributed},
the ADMM routine for solving problem \eqref{eq:prob_admm} is
\begin{subequations}
	\begin{align}
		\balp^{k+1} & = \arg \min_{ \bone^\top \balp = \eta } L_\rho(\balp,\bbeta^k,\blam^k),
		\label{eq:admm_alp}
		\\
		\bbeta^{k+1} & = \arg \min_{ \bbeta > \bzero } L_\rho(\balp^{k+1},\bbeta,\blam^k),
		\label{eq:admm_beta}
		\\
		\blam^{k+1} & = \blam^{k} + \rho ( \balp^{k+1} - \bbeta^{k+1} ),
	\end{align}
\end{subequations}
for $k= 0, 1,\ldots$ and given a starting point $( \balp^0,\bbeta^0,\blam^0 )$.

The update step in \eqref{eq:admm_beta} is handled as follows.
The problem in \eqref{eq:admm_beta} collapses into a number of scalar problems
\beq \label{eq:beta_probs}
\beta_i^{k+1} = \arg \min_{\beta > 0} \, h(\beta) - \lambda_i^k \beta + \frac{\rho}{2} ( \alpha_i^{k+1} - \beta )^2,
\eeq
for  $i=1,\ldots,N$.
The solution to  problem \eqref{eq:beta_probs} can be obtained by finding $\beta > 0$ such that the derivative of the objective function of \eqref{eq:beta_probs} is zero; i.e.,
\beq \label{eq:find_sol}
0 = h'(\beta) - \lambda_i^k - \rho ( \alpha_i^{k+1} - \beta ).
\eeq
The derivative $h'$ does not admit a simple expression, although it is computable. It can be shown that $h'(\beta) = (\beta- 1) \psi'(\beta)$, where $\psi'$ is the trigamma function.
The trigamma function does not have a closed form, but major numerical software (such as MATLAB) has specialized routines for computing the value of $\psi'(\beta)$ of a given $\beta$.
We numerically find the solution to \eqref{eq:find_sol} by line search, specifically, the bisection search.

The update step in \eqref{eq:admm_alp} is a convex quadratic program with one linear equality constraint, and it has a closed form.
To formulate properly, rewrite $g$ in \eqref{eq:fn_g}  as a quadratic function
\ifconfver
\begin{align*}
	g(\balp) & = - \balp^\top \underbrace{\left[ \tfrac{1}{\sigma^2 \eta} \left( \bA^\top \by - \tfrac{1}{2(\eta+1)} \diag(\bA^\top \bA)  \right)    \right]}_{:= \bb} +  \\
	& + \tfrac{1}{2} \balp^\top \underbrace{\left( \tfrac{1}{\sigma^2(\eta+1)\eta} \bA^\top \bA \right)}_{:= \bC} \balp,
\end{align*}
\else
\begin{align*}
	g(\balp) & = - \balp^\top \underbrace{\left[ \tfrac{1}{\sigma^2 \eta} \left( \bA^\top \by - \tfrac{1}{2(\eta+1)} \diag(\bA^\top \bA)  \right)    \right]}_{:= \bb}  + \tfrac{1}{2} \balp^\top \underbrace{\left( \tfrac{1}{ \sigma^2 (\eta+1)\eta} \bA^\top \bA \right)}_{:= \bC} \balp,
\end{align*}
\fi
where $\diag(\bX)= (x_{11},\ldots,x_{nn})$.
The problem in \eqref{eq:admm_alp} can be written as
\beq \label{eq:admm_alp_prob1}
\min_{ \bone^\top \balp = \eta } \, - \balp^\top ( \bb + \rho \bbeta^k - \blam^k ) + \tfrac{1}{2} \balp^\top ( \bC + \rho \bI) \balp.
\eeq
To derive a closed-form solution to problem \eqref{eq:admm_alp_prob1}, note that any $\balp$ satisfying $\bone^\top \balp = \eta$ can be equivalently represented by
\beq \label{eq:alp_zeta}
\balp = \tfrac{\eta}{N} \bone + \bU \bzeta, \quad  \bzeta \in \Rbb^{N-1},
\eeq
where $\bU \in \Rbb^{N \times (N-1)}$ is a semi-orthogonal matrix such that $\bU^\top \bone = \bzero$.
By the change of variables in \eqref{eq:alp_zeta}, we can rewrite problem \eqref{eq:admm_alp_prob1} as
\beq \label{eq:admm_alp_prob2}
\min_{  \bzeta \in \Rbb^{N-1} } \, - \bzeta^\top \bU^\top ( \bb + \rho \bbeta^k - \blam^k - \tfrac{\eta}{N} \bC \bone  ) + \tfrac{1}{2} \bzeta^\top \bU^\top ( \bC + \rho \bI) \bU \bzeta.
\nonumber
\eeq
The above problem, as an unconstrained quadratic program, has a closed-form solution
\[
\bzeta^{k+1} = ( \bU^\top \bC \bU + \rho \bI )^{-1}  \bU^\top ( \bb + \rho \bbeta^k - \blam^k - \tfrac{\eta}{N} \bC \bone  );
\]
and we obtain $\balp^{k+1}$ by $\balp^{k+1}= \tfrac{\eta}{N} \bone + \bU \bzeta^{k+1}$.

\bibliographystyle{IEEEtran}
\bibliography{refs}

\end{document}